\documentclass[aps,pre,amsmath,amssymb,twocolumn,groupedaddress]{revtex4-1}

\usepackage[utf8]{inputenc}
\usepackage{graphicx}
\usepackage{dcolumn}
\usepackage{bm}
\usepackage{amssymb}
\usepackage{amsbsy}
\usepackage{color}
\usepackage{amscd}
\usepackage{mathrsfs}
\usepackage{gensymb} 

\usepackage{mathrsfs}
\usepackage{calligra}
\DeclareMathAlphabet{\mathcalligra}{T1}{calligra}{m}{n}

\usepackage{hyperref}

\makeatletter
\newcommand*{\rom}[1]{\expandafter\@slowromancap\romannumeral #1@}
\makeatother

\begin{document}

\title{Diffusiophoresis driven colloidal manipulation and shortcuts to adiabaticity}

\author{Parvin Bayati}
\affiliation{Universit\'{e} Paris-Saclay, CNRS, LPTMS, 91405, Orsay, France} 

\author{Emmanuel Trizac}
\affiliation{Universit\'{e} Paris-Saclay, CNRS, LPTMS, 91405, Orsay, France}

\begin{abstract}
While compressing a colloidal state by optical means alone has been previously achieved through a specific time-dependence of the trap stiffness, realizing quickly the reverse transformation stumbles upon the necessity of a transiently expulsive trap. To circumvent
this difficulty, we propose to drive the colloids
by a combination of optical trapping and diffusiophoretic
forces, both time-dependent. Forcing via diffusiophoresis is enforced by controlling the salt concentration at
the boundary of the domain where the colloids are confined. The method takes advantage of the 
separation of time scales between salt and colloidal dynamics, and realizes a fast decompression in an optical trap that remains
confining at all times. We thereby obtain a so-called shortcut to adiabaticity protocol where
colloidal dynamics, enslaved to salt dynamics, can nevertheless be controlled as desired.
\end{abstract}

%
\maketitle


\section{Introduction\label{sec:intro}}

Shortcut to adiabaticity protocols have been introduced 
as fast alternatives to otherwise time consuming transformations, where the parameters controlling a system are slowly (adiabatically) changed~\cite{torrontegui2013shortcuts,guery2019shortcuts}.
Applied to quantum systems, they allow for manipulation on timescales shorter than the decoherence time,
with applications in metrology and interferometry~\cite{guery2019shortcuts}. They are also useful when applied to classical devices, providing for instance a better control of micromechanical oscillators, with applications 
in faster atomic force imaging techniques~\cite{le2016fast}. 
Shortcut to adiabaticity protocols have consequently been studied and tested successfully in different 
systems~\cite{chen2010fast,walther2012controlling,deng2013boosting,guery2014nonequilibrium,
rossnagel2014nanoscale,chen2015fast,dechant2017underdamped,impens2017shortcut,patra2017shortcuts,abah2018performance,funo2020shortcuts,baldassari2020}. 
In these approaches, one or two control parameters are engineered such that the equilibrium is reached in a chosen 
short time, using processes which are not necessarily heat-exchange-free nor isothermal. A beneficial feature of these methods lies in the multiplicity 
of admissible protocols, which allows to select those
with desirable properties, in terms of robustness or optimality~\cite{aurell2011optimal,guery2019shortcuts,zhang2020work}.
It is worth stressing that we use here the word ``adiabaticity'' in the quantum mechanics inspired meaning of ``slow enough''; it should thus not be confused with the thermodynamic meaning of being heat--exchange free~\cite{plata2020finite}.

The present work is devoted to manipulating trapped colloidal systems. It is related to the Engineered Swift Equilibration (ESE) protocol presented in~\cite{martinez2016engineered}.
Following this framework, it will be here assumed that the colloidal dynamics is overdamped, which proves to be an
experimentally very accurate approximation~\cite{plata2020finite}.
By appropriately tuning the trap stiffness as a function of time, a fast compression of a colloidal state was achieved, with a close to 100-fold acceleration compared to the bare relaxation time~\cite{martinez2016engineered}. 
During the process, the probability distribution function of the colloid position is kept Gaussian while the trapping potential remains quadratic, facilitating reaching the equilibrium at the final time. 
Generalizations to the underdamped regime, and to two hydrodynamically interacting colloids 
have been worked out in Refs.~\cite{chupeau2018engineered,dago2020engineered}. Besides, variants have been proposed allowing for non-conservative forces~\cite{li2017shortcuts}, or transitions between non-equilibrium steady states
\cite{baldassari2020}.

A drawback of ESE protocols appears when comparing compression and expansion
of systems involving confinement with optical tweezers. A state can be compressed
by adequately choosing the time dependence of the intensity of the trapping laser; increasing the speed of the transformation simply requires
a larger laser intensity range~\cite{martinez2016engineered}. On the other hand, expanding a trapped system in a quick fashion requires to create, transiently, an expulsive rather than 
confining potential~\cite{martinez2016engineered,chupeau2018engineered}. This 
is necessary to guarantee a fast displacement of the colloids
towards their statistically relevant end position. In other words, the trap stiffness not only needs to follow a precise
time dependence, but also to be negative during some time window. 
This is a difficult experimental challenge, for instance when trapping is realized with optical tweezers~\cite{albay2020realization}. 
A partial solution to this shortcoming was proposed with the method worked out
in Ref.~\cite{plata2019optimal}, where protocols conditioned
to using non-negative stiffness have been derived, optimizing furthermore
the (mean) work performed by the operator. 
Here, we shall explore a
different venue for effectively achieving transient negative stiffness.
Our proposition is an alternative to the thermal bath engineering strategy put forward in Ref.~\cite{chupeau2018thermal}. There, an externally 
controlled noise is applied to the center of the trap, mimicking a time-dependent effective temperature that necessarily exceeds that of the bath in which the system is immersed (say water).
The possibility to decompress the system under study without 
repulsive potentials ensues: advantage is taken of an 
enhanced diffusion. Yet, this technique does not suit for ESE compression~\cite{comment55}. On the contrary, the present 
combination of optical and phoretic forces is operational 
not only to allow for decompression, but can be as well transposed
to compression.
Our approach also differs from the optical feedback trap realizations proposed in~\cite{albay2020realization},
in so far as we do not operate any retro-action on the system,
but proceed in a purely feed-forward manner.

Since achieving fast decompression ESE paves the way towards possible application in designing or 
improving the performance of small-scale heat engines~\cite{rossnagel2016single,erdman2019maximum,hartmann2020many,plata2020building} 
and thermodynamic cycles in Brownian and active 
systems~\cite{schmiedl2008efficiency,blickle2012realization,martinez2016brownian,plata2020building,rosales2020optimal,ekeh2020thermodynamic,kumari2020stochastic,holubec2020active},
we search for operational procedures of augmented ESE, to circumvent the decompression bottleneck.
We propose here a protocol which takes advantage of the diffusiophoresis of colloids.
Diffusiophoresis refers to the migration of colloids and macro--molecules immersed in a solution under a 
gradient of solute, that is driven by a direct surface–solute interaction~\cite{anderson1989colloid,prieve1984motion}.
Recently, diffusiophoretic forces emerged in various microfluidic applications~\cite{zhao2012advances}, 
for enhancing transport rate of flow~\cite{abecassis2008boosting} and designing novel types of active 
microswimmers~\cite{paxton2004catalytic,bayati2016dynamics,bayati2019electrophoresis}, as well as for industrial applications such as underground oil and gas recovery~\cite{sheng2014critical}.
Other applications include the effective and long-lasting 
cleaning or removal of particles and droplets from deep pores~\cite{ault2017diffusiophoresis,shin2018cleaning}, 
and enhanced oil recovery from deep wells~\cite{katende2019critical,myint2015thin,marbach2019osmosis}.
A particular feature of diffusiophoresis emerges in electrolytes (i.e., ionic solutions where the solute is a salt) where
the ions and the particle surface interact through electrostatic forces: the colloidal (mean) velocity becomes proportional to the gradient of the logarithm 
of the solute concentration, i.e., ${\bf V}_{\mathrm{DP}} \propto \nabla \ln C$, 
at variance with neutral solutions where this dependency is linear in the concentration gradient~\cite{prieve1984motion}. 
This feature allows for efficient driving of colloids~\cite{palacci2010colloidal,shin2016size},
and guarantees significant forcing even in regime
of small solute concentrations.
It also opens the possibility of trapping colloids and 
macro--molecules via osmotic forces, as recently  achieved~\cite{palacci2012osmotic,shi2016diffusiophoretic}.

In this paper, we propose an ESE strategy that takes advantage of diffusiophoresis.
As the phoretic force direction depends on both the solute gradient direction and the solute--surface interaction, 
creating a repulsive potential is achievable
by appropriately tuning the solute concentration on the boundaries of the system as a function
of time~\cite{ha2019dynamic,shi2016diffusiophoretic}. Accordingly, our 
approach allows for both an accelerated expansion or an accelerated compression of a colloid state in harmonic or non-harmonic confinement.
Our analysis restricts to low density systems, where colloidal dynamics does
not result from colloid-colloid interactions, but from other types of forces.
This logic can be pushed further, considering that a single colloid is being manipulated,
in a repeated fashion in order to gather statistics and generate the 
distribution functions that will be object of interest here. So was the situation in the experiment reported in Ref.~\cite{martinez2016engineered}.  

The rest of the paper organizes into six sections. We address the basics of ESE protocols in the overdamped regime 
and present its governing equations in section~\ref{sec:ESE}.
Section~\ref{sec:DP} is devoted to the key equations of 
diffusiophoresis.
Then, we combine the ESE method with diffusiophoresis to introduce the 
diffusiophoresis driven protocol in section~\ref{sec:ESE_DP_G} for Gaussian colloidal density and harmonic trapping potentials.
In section~\ref{sec:res_1D}, we discuss the consistency of these results.
We consider non-Gaussian states and non-harmonic potentials in section~\ref{sec:ESE_DP_non_G_pdf}, while concluding remarks are presented in
section~\ref{sec:conclusion}.

\section{ESE for Gaussian distributions\label{sec:ESE}} 
\begin{figure}[!ht]
\centering \includegraphics[width=0.9\linewidth]{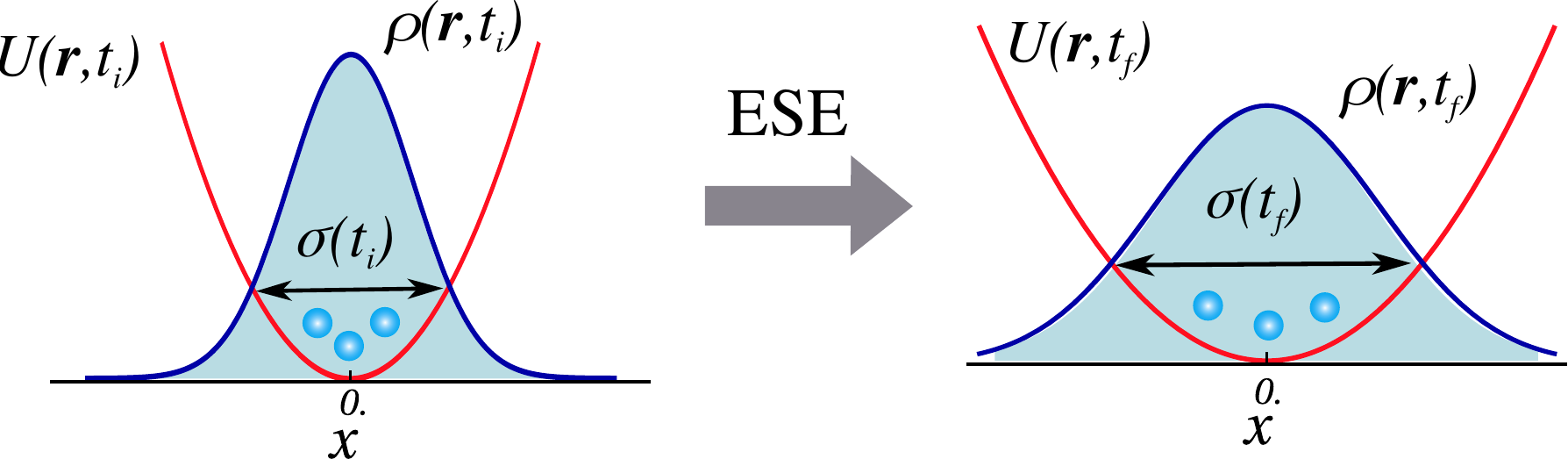}
\caption{\label{fig:model} Schematic decompression of a colloidal state via ESE
where $x$ denotes a Cartesian coordinate of the position $\mathbf{r}$. 
At initial time $t_{\mathrm{i}}$, the colloids are at equilibrium with a Gaussian distribution $\rho({\bf r},t_{\mathrm{i}})$ having variance
$\sigma_{\mathrm{i}}^2 = k_{\mathrm{B}} T /\kappa_{\mathrm{i}}$, trapped in a harmonic potential $U({\bf r},t_{\mathrm{i}})$. The aim is to find a suitable 
scheme to expand the system to a final equilibrium state with variance $\sigma_{\mathrm{f}}^2 = k_{\mathrm{B}} T /\kappa_{\mathrm{f}}$,
in a time $t_{\mathrm{f}}$ shorter than the relaxation time $\tau_{\mathrm{relax}}$. Since the whole analysis is performed at single colloid level, it is assumed that the concentration of colloids is small enough
so that inter-colloidal interactions are negligible. Alternatively, one can envision all transformations as pertaining to a single trapped colloid, with an experiment repeated a large number of times to gather statistics~\cite{martinez2016engineered,dago2020engineered}; in doing this, the colloid initial position, when the transformation begins, samples the distribution $\rho(\mathbf{r},t_{\mathrm{i}})$.}
\end{figure}

The system under study is a (overdamped) Brownian colloidal particle in water, trapped in a harmonic potential $U =  \kappa r^2 / 2$, where $r$ is the distance to the trap center. This external potential is
usually established by optical tweezers. 
We seek for a transition where the stiffness $\kappa$ is changed from an initial value $\kappa_{\mathrm{i}}$ to a 
final value $\kappa_{\mathrm{f}}$ in a given time $t_{\mathrm{f}}$. 
The idea is to find a suitable function for the stiffness $\kappa(t)$ 
such that, as demonstrated in Fig.~\ref{fig:model},
the system evolves from an equilibrium state with initial variance $\sigma_{\mathrm{i}}^2 = k_{\mathrm{B}} T/\kappa_{\mathrm{i}}$ to the final value $\sigma_{\mathrm{f}}^2 = k_{\mathrm{B}} T/\kappa_{\mathrm{f}}$. This should be achieved
faster than the natural relaxation time 
$ \tau_{\mathrm{relax}} = \gamma / \kappa_{\mathrm{f}}$,
that would be required to reach equilibrium  after a sudden change of stiffness from $\kappa_{\mathrm{i}}$ to $\kappa_{\mathrm{f}}$.
Here, 
$\gamma$ is the colloid drag coefficient, related to the diffusion 
coefficient of the colloid $D_{\mathrm{c}}$ through $\gamma = k_{\mathrm{B}} T /D_{\mathrm{c}}$, where $k_{\mathrm{B}}$ and $T$ represent the Boltzmann constant and 
the water bath temperature, respectively. 
The system is in equilibrium at initial and final times, but not during the process. 
The bath temperature is constant.  
Note that with the above relations, the 
relaxation time can also be written 
$\tau_{\mathrm{relax}} = \sigma_{\mathrm{f}}^2/D_{\mathrm{c}}$: it corresponds to the time 
needed to explore diffusively a region of space of 
linear size $\sigma_{\mathrm{f}}$.

\subsection{The formalism}

The overdamped Langevin equation describes the dynamics of the colloid at position ${\bf r}$ relative to the trap center at position $\mathbf{0}$, as
\begin{equation}
\gamma \dot{\bf r} = -\nabla U({\bf r},t) + {\bf \xi}(t) = - \kappa(t) ~{\bf r} + {\bf \xi}(t),
\label{eq:langevin}
\end{equation}
where ${\bf \xi}(t)$ is a random force, a white noise with 0 mean and correlation $\langle\xi_{\mathrm{i}}(t)\xi_j(t')\rangle=2 \gamma k_{\mathrm{B}} T \delta_{ij}\delta(t-t')$.
This very treatment, neglecting colloidal-colloidal interactions, 
thus holds for dilute enough systems, a situation often met experimentally.
Accordingly, the colloidal density $\rho({\bf r},t)$ is governed by
the Fokker-Planck equation:
\begin{equation}
 \partial_t \rho({\bf r},t) = \nabla \cdot \left(D_{\mathrm{c}} \nabla \rho({\bf r},t) + \frac{1}{\gamma} \nabla U \rho({\bf r},t)\right).
 \label{eq:fokker}
\end{equation}
Control is achieved here by 
changing of the stiffness $\kappa(t)$ such that
the density remains Gaussian at all times
\begin{equation}
 \rho ({\bf r},t) = \left(\frac{\alpha(t)}{\pi}\right)^{n/2} \exp(-\alpha(t) r^2),
 \label{eq:G_pdf}
\end{equation}
where $n$ refers to the system's dimension. The inverse variance $\alpha(t)$ can be chosen arbitrarily, fulfilling the equilibrium boundary conditions
$\alpha(0)=\kappa_{\mathrm{i}}/(2 k_{\mathrm{B}} T), \alpha(t_{\mathrm{f}})=\kappa_{\mathrm{f}}/(2 k_{\mathrm{B}} T)$,
to which we add the smoothness conditions
$\dot{\alpha}(0)=\dot{\alpha}(t_{\mathrm{f}})=0$,
experimentally more friendly than discontinuities (which are, strictly speaking,
possible and do indeed arise when optimal features are sought~\cite{schmiedl2007optimal,aurell2011optimal,plata2019optimal}).
Non--Gaussian distributions will be addressed in section~\ref{sec:ESE_DP_non_G_pdf}.
We do not need to specify the colloid's radius, which enters implicitly in the 
friction coefficient $\gamma$. 

In order to obtain the time-dependent forcing $\kappa(t)$, 
we follow Ref. \cite{martinez2016engineered} and substitute Eq.~\eqref{eq:G_pdf} into the Fokker-Planck equation~\eqref{eq:fokker},
and obtain:
\begin{equation}
\frac{n \dot{\alpha}(t)}{2 \alpha(t)} - \dot{\alpha}(t) r^2 =  
\frac{\kappa(t)}{\gamma} (n-2 \alpha(t) r^2) - 2 \alpha(t) D_{\mathrm{c}}(n -2\alpha(t) r^2),
\end{equation}
which holds for any position $r$. 
This implies that
\begin{equation}
 \kappa(t) = \frac{\dot{\alpha}(t)}{2 \alpha(t)} \gamma + 2 k_{\mathrm{B}} T \alpha(t),
 \label{eq:kappa}
\end{equation}
irrespective of dimension $n$. This indicates 
how to tune the trapping laser intensity with time. This led to an operational protocol in Ref.~\cite{martinez2016engineered},
whereby a colloidal state was compressed 
with a close to 100-fold gain in the 
time required to reach the desired equilibrium. 
%

Before discussing the difficulty met when 
transposing this idea from compression to decompression,
we point out the relevant dimensionless parameter quantifying acceleration. It is given by the ratio of time scales 
${\cal A} = \tau_{\mathrm{relax}} / {t_{\mathrm{f}}} = \sigma_{\mathrm{f}}^2 / (D_{\mathrm{c}} t_{\mathrm{f}})$
and measures the desired acceleration factor.
Noting that $\dot{\alpha}$ has to be negative for a decompression processes, it follows from Eq.~\eqref{eq:kappa} that
for sufficiently fast decompressions (${\cal A} \gg 1$), 
the potential should become transiently ``expulsive'' ($\kappa<0$), 
irrespective of the choice of $\alpha(t)$.
While feedback-based protocols have been
proposed to this aim~\cite{albay2020realization}, achieving 
time-dependent and transiently repulsive confinements remains  
experimentally problematic, which motivates 
us to propose an alternative.

\section{Diffusiophoresis for manipulating colloids\label{sec:DP}}
%
Diffusiophoresis is an out of equilibrium phenomenon by which colloids immersed in a solution migrate in 
response to a gradient of the solute concentration $\nabla C({\bf r},t)$.
The solute gradient combined with direct interaction of particle--solute
results in an osmotic pressure gradient in the thin interacting layer around the colloid. This pressure gradient creates a 
fluid flow and, consequently, colloid motion with velocities proportional to the solute gradient%
~\cite{anderson1989colloid,prieve1984motion}.
Moreover, an interesting feature occurs in the case of electrolytes (a salt as a solute) and charged colloids, when 
the salt-colloid interaction potential has electrostatic nature. In this particular case, the thickness of the interacting layer (diffuse Debye layer)
is inversely proportional to the ions concentration. 
This leads to  
a colloids' velocity that is proportional to the gradient of logarithm of the ions concentration~\cite{anderson1989colloid,prieve1984motion}
\begin{equation}
\label{v_DP}
{\bf V}_{\mathrm{DP}} = \Gamma_{\mathrm{c}} \nabla \ln C({\bf r},t),
\end{equation}
where $\Gamma_{\mathrm{c}}$ is the phoretic mobility of colloidal particles. 
This quantity is related to the 
interaction potential between colloids and salt, 
and can be positive or negative~\cite{note10}.
Accordingly, the diffusiophoretic force exerted on the colloid and the corresponding potential energy
have the same log-dependency~\cite{comment202},
\begin{equation}
\label{eq:UDF}
{\bf F}_{\mathrm{DP}} = \gamma \Gamma_{\mathrm{c}} \nabla \ln C({\bf r},t), 
~~~~ U_{\mathrm{DP}} = -\gamma \Gamma_{\mathrm{c}} \ln C({\bf r},t).
\end{equation}

\begin{figure}[!ht]
\centering \includegraphics[width=0.9\linewidth]{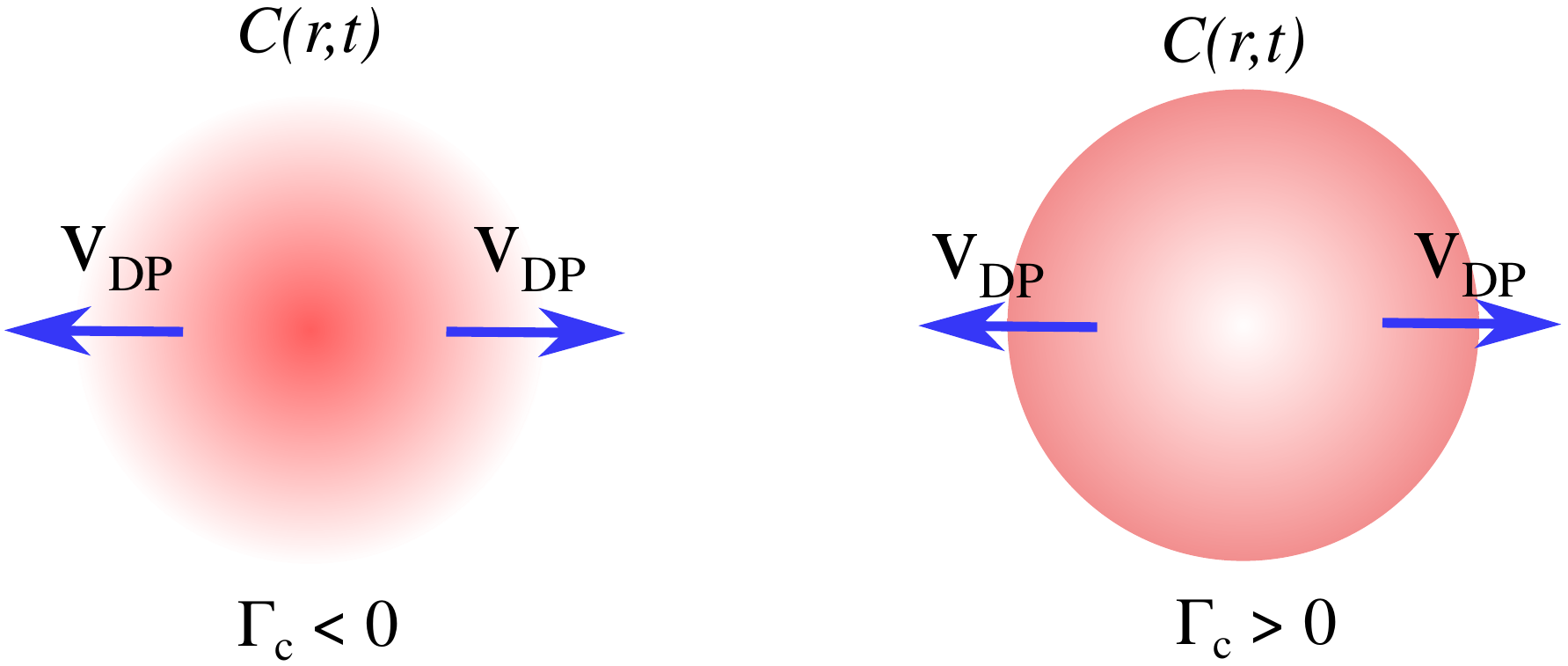}
\caption{\label{fig:con_model} The gradient of ions concentration that should be applied to the system to establish a repulsive diffusiophoretic force required 
for a quick decompression. Depending on the sign of the phoretic mobility $\Gamma_{\mathrm{c}}$, the gradient should exhibit larger concentrations at the center ($\Gamma_{\mathrm{c}}<0$) or at the boundaries ($\Gamma_{\mathrm{c}}>0$).}
\end{figure}

According to this relation, the direction and strength of ${\bf F}_{\mathrm{DP}}$ can be tuned by the concentration
gradient and the phoretic mobility $\Gamma_{\mathrm{c}}$.
This feature yields the possibility of creating a repulsive force required for an ESE decompression protocol.
Therefore, to make the colloids migrate away from the center of the system (i.e., to realize an expansion),
a concentration gradient in the direction towards (respectively away from) the center should be applied for the case of negative 
(respectively positive) phoretic mobilities. 
It means that, as depicted in Fig.~\ref{fig:con_model}, the concentration on the boundaries $\phi(t)$ should be decreased 
(with respect to the concentration at the center) for the case of negative phoretic mobilities, and be increased
for positive phoretic mobilities.
Interestingly, thanks to the log-sensing, when time and position dependencies of the salt concentration $C({\bf r},t)$ factorize,
$\nabla \ln C = \nabla C/C$ becomes independent of time. This happens in the long time limit 
when salt diffuses out of the system, so that $C\to 0$ ~\cite{palacci2010colloidal,palacci2012osmotic},
with the surprising consequence that the force
acting on the colloid is maintained while
the solute at its origin fades away.

To complete the diffusiophoresis discussion, we note that
for small colloidal charges (Debye-H{\"u}ckel approximation) and in the thin Debye layer limit 
(i.e., with extension much smaller than colloid size),
the concentration of positive and negative ions outside the Debye layer are approximately equal 
and governed by the diffusion equation~\cite{prieve1984motion}
\begin{equation}
 \partial_t C({\bf r},t) = D_{\mathrm{s}} \nabla^2 C({\bf r},t).
 \label{eq:diffusion}
\end{equation}
This is supplemented 
with initial and boundary conditions as follows:
\begin{equation}
C({\bf r},0) = C_0,~~~~~~ C({\bf r},t)|_{B} = \phi(t),
 \label{eq:IBC}
\end{equation}
where
$C_0$ is the initial uniform ions concentration. Here, $B$ refers to the boundaries of the system. For simplicity, we
assume that the diffusion coefficient of positive and negative ions are equal and given by $D_{\mathrm{s}}$~\cite{note10}.

Finally, we note that the simplest way to establish a concentration gradient is to set a 
vanishing concentration $\phi(t)$ such that $\phi(t\geq	0)=0$.
Any choice of $\phi(t)$ that verifies $\phi=0$ for $t\geq	 0$ is possible. For example, 
one option is setting $\phi(t)$ \textit{suddenly} to zero at $t=0$
from a given value $C_0$.
As alluded to above, this leads after a
salt-characteristic relaxation time ($\tau_{\mathrm{s}}$) to 
a time-position factorized salt concentration, 
with an associated time-independent forcing,
see~\ref{appen:expa_beyond_ESE}.
However, although the resulting procedure is admissible for accelerating decompression, the final state is not controllable: the resulting stiffness 
is system size dependent, but cannot be tuned, as discussed in~\ref{appen:expa_beyond_ESE} for a 1D system.
Moreover, the configuration under study establishes
a salt gradient towards the center $\mathbf{r=0}$. Hence, as sketched in Fig.~\ref{fig:con_model}, this method can be used to decompress colloids with
negative phoretic mobility (and compress colloids with positive phoretic mobility), but
it is not a versatile approach for shortcutting adiabaticity, and we search for another route, where $\phi(t)$ will be deduced rather than imposed.

\section{Diffusiophoresis driven ESE: Gaussian states}
\label{sec:ESE_DP_G}

We next combine optical and diffusiophoretic drivings to establish our ESE protocol. The colloidal system is subject to optical confinement that can be time-dependent, and is furthermore in an electrolyte 
that can be controlled through the boundaries
of the system, as performed in~\cite{palacci2012osmotic,ha2019dynamic}.
It might be thought that such a surface 
driving through the boundary concentration $\phi(t)$ does not provide enough command over
the system, since salt invariably diffuses
in the bulk of the solution, but we will see below that the present setting nevertheless offers
interesting means to shape shortcutting protocols.

The total potential acting on the colloid is composed of a harmonic optical and a diffusiophoretic contribution, as
\begin{equation}
\label{eq:U_x2}
U= U_{\mathrm{op}} + U_{\mathrm{DP}} = \frac{1}{2} \kappa_{\mathrm{op}}(t) r^2 -\gamma \Gamma_{\mathrm{c}} \ln C({\bf r},t).
\end{equation}
Both contributions are time dependent; in line with
our underpinnings, we set the constraint 
$\kappa_{\mathrm{op}}>0$ for the optical contribution.
In addition, we assume isotropy (so that 
$C$ only depends on $r=|{\bf r}|$) and analyticity.
The concentration, extremum at ${\bf r}={\bf 0}$, can then be expanded at small $r$
to yield
\begin{eqnarray}
\label{eq:U_x2_appro}
U &&= \frac{1}{2}\left(\kappa_{\mathrm{op}}(t)
-\gamma \Gamma_{\mathrm{c}} \partial^2_{r}  \ln C(r,t)\biggl|_{r=0} \right) r^2 + {\cal O}(r^4)\nonumber\\
&& \simeq\frac{1}{2} \kappa(t) r^2 ,
\end{eqnarray}
(up to a position-independent shift in the potential origin, which is not relevant here) with the diffusiophoretic contribution to stiffness 
\begin{equation}
 \kappa_{\mathrm{DP}}(t)  \equiv \kappa(t)-\kappa_{\mathrm{op}}(t) = 
 -\gamma \Gamma_{\mathrm{c}} \partial^2_{r}  \ln C(r,t)|_{r=0}.
 \label{eq:kappa_DP}
\end{equation}
We demand that the total stiffness $\kappa(t)$ follows the rule given by Eq.~\eqref{eq:kappa},  for a chosen $\alpha(t)$. While the sum $\kappa_{\mathrm{DP}} + \kappa_{\mathrm{op}}$ is hence known, the choice of
 $\kappa_{\mathrm{op}}$ enjoys some flexibility, and
has to be specified in a convenient way. 
The simplest choice is arguably $\kappa_{\mathrm{op}}=\kappa_{\mathrm{i}}$, the only temporal dependant term in the total stiffness being borne by $\kappa_{\mathrm{DP}}(t)$.
Obtaining a steady non vanishing value for $\kappa_{\mathrm{DP}}$ requires that $\phi(t>t_{\mathrm{f}}) =0$: otherwise, salt diffuses to reach a uniform concentration across the system,
which cancels out the diffusiophoretic force. 
However, demanding {\it a priori} a vanishing $\phi(t>t_{\mathrm{f}})$ is a somewhat nontrivial constraint (see below). We will follow a different strategy where 
the signal $\phi(t)$ is the end product of the calculation, rather than an input;
we will rather use the freedom of having a
time-dependent optical potential, such that
the diffusiophoresis provides a negative total stiffness (repulsive force), and the optical tweezers tune the initial and 
final states.

From Eq.~\eqref{eq:kappa} and the diffusion equation~\eqref{eq:diffusion} which relates 
$\partial^2_{r}$ to $\partial_t$,
we get the following constraint for the concentration at the center 
(see~\ref{appen:C0} for details)
\begin{equation}
 -\frac{\gamma \Gamma_{\mathrm{c}}}{n~D_{\mathrm{s}}}  \partial_t\ln C(0,t)
 = \frac{\dot{\alpha}(t)}{2 \alpha(t)} \gamma  + 2 k_{\mathrm{B}} T\alpha(t) - \kappa_{\mathrm{op}}(t).
\label{eq:den_cond}
\end{equation}
Solving for $C(0,t)$ yields
\begin{eqnarray}
&&C(0,t) =\nonumber\\
&&C_0 \left(\frac{\alpha}{\alpha_{\mathrm{i}}}\right)^
{-\frac{n D_{\mathrm{s}}}{2\Gamma_{\mathrm{c}}}}
 \exp\left(-\frac{n D_{\mathrm{s}}}{\gamma \Gamma_{\mathrm{c}}} \int_{0}^{t} (2k_{\mathrm{B}} T\alpha(\tau)-\kappa_{\mathrm{op}}(\tau))d\tau\right).\nonumber\\
\label{eq:C_0}
\end{eqnarray}
Finally, the concentration $\phi(t)$ at the boundary
($r=R$), is extracted from a Taylor expansion of the density at $r=0$:
\begin{eqnarray}
&&\phi(t) = C(r=R,t)= \sum_{j=0}^{\infty} \frac{1}{j!} R^\mathrm{j} \partial^\mathrm{j}_r C(r,t)|_{r=0} \nonumber\\
&&= \sum_{j=0}^{\infty} \frac{1}{(2j)!} R^\mathrm{2j} \partial^\mathrm{2j}_r C({\bf r},t)|_{r=0}\nonumber\\
&&=\sum_{j=0}^{\infty} \frac{ R^\mathrm{2j}}{(2j)! D_{\mathrm{s}}^\mathrm{j}} 
\left(\prod_{m=1}^j\frac{2m-1}{2m-2+n}\right) \partial_t^\mathrm{j} C(0,t).
\label{eq:phi_Taylor}
\end{eqnarray}
Here, we have made use of
\begin{equation}
\partial^\mathrm{2j}_r C({\bf r},t)|_{r=0} =  
\left(\prod_{m=1}^j\frac{2m-1}{2m-2+n}\right) \frac{ 1}{D_{\mathrm{s}}^\mathrm{j}} \partial_t^\mathrm{j} C(0,t),
\end{equation}
which is obtained by taking successive derivatives of the diffusion equation \eqref{eq:diffusion} and invoking 
$\partial^j_{r} C |_{ r=0} =0$ (for $j$ odd).

To summarize the procedure, the operator starts by the choice of $\alpha(t)$, which is proportional to the inverse variance of colloidal density (more precisely, $\sigma^2 = (2\alpha)^{-1}$). Then, a suitable function for the optical stiffness $\kappa_{\mathrm{op}}(t)$ should be chosen, see the next section.
Next, by calculating $C(0,t)$ from Eq.~\eqref{eq:C_0}, we subsequently obtain $\phi(t)$ from Eq.~\eqref{eq:phi_Taylor}, truncating the infinite summation appropriately.
There is at this point no guarantee that 
$\phi(t)$ and, consequently, $C({\bf r},t)$ be positive. Working for the parameter range where both functions are non-negative sets a limitation to the 
approach, that will be assessed.
To this end, we will provide in the next section a phase portrait for $\phi(t)$ as a function of the acceleration factor 
${\cal A}$ ($=\tau_{\mathrm{relax}}/t_{\mathrm{f}}$)
and the ratio of initial and final stiffnesses $\kappa_{\mathrm{i}}/\kappa_{\mathrm{f}}$.

There are three time-scales in our problem: 
the relaxation time of the colloidal particle $\tau_{\mathrm{relax}}= \sigma_{\mathrm{f}}^2/D_{\mathrm{c}}$, 
the operation time $t_{\mathrm{f}}$, and the salt diffusion time
 $\tau_{\mathrm{s}} = R^2/ D_{\mathrm{s}}$ (where $R$ denotes the system size, see also~\ref{appen:expa_beyond_ESE}). 
Due to the size asymmetry, the diffusion coefficient of ions
is much larger than that of colloidal particles
($D_{\mathrm{s}} \gg D_{\mathrm{c}}$).
Moreover, we cannot suppress the Brownian diffusion of ions (salt). Therefore, 
the final time $t_{\mathrm{f}}$ should be larger than $\tau_{\mathrm{s}}$ in order for the diffusion of ions not to perturb the final state reached at 
$t_{\mathrm{f}}$: any remaining non equilibrium ionic dynamics
for $t>\tau_{\mathrm{s}}$ would impinge on the colloidal state and make it non stationary. Therefore, we impose that
$\tau_{\mathrm{s}} < t_{\mathrm{f}} \ll \tau_{\mathrm{relax}}$, which means that 
$R^2/D_{\mathrm{s}} \ll \sigma_{\mathrm{f}}^2/D_{\mathrm{c}}$. This sets an upper 
bound for the system size $R$, which controls the
time scale for salt dynamics, and should be kept
low for our approach to be operational~\cite{comment203}.

As a final remark, 
we stress that we have cut the infinite Taylor expansion Eq.~\eqref{eq:phi_Taylor}. To assess the possible resulting loss 
of precision for the whole scheme,
we shall test the consistency of the results.
To this end, having the boundary condition for salt concentration,
$\phi(t)$, we solve the diffusion equation~\eqref{eq:diffusion} numerically and obtain $C({\bf r},t)$. We then check that
$\phi(t)$ is reasonably close to $C({\bf r}={\bf R},t)$. Next, having $C({\bf r},t)$, 
we can calculate stiffnesses ($\kappa(t)$ and $\kappa_{\mathrm{DP}}(t)$) and compare them with the target functions, as given by Eqs.~\eqref{eq:kappa} and \eqref{eq:kappa_DP}. Finally, we need to evaluate 
the final variance $\sigma_{\mathrm{f}}^2 = \int r^2 \rho({\bf r},t_{\mathrm{f}})~d^n{\bf r}$ which
should be equal to $k_{\mathrm{B}} T/\kappa_{\mathrm{f}}$.

\section{Results in one dimension: Gaussian states}
\label{sec:res_1D}

\subsection{Characterizing the acceptable protocols}
\begin{figure}[!ht]
\centering \includegraphics[width=0.85\linewidth]{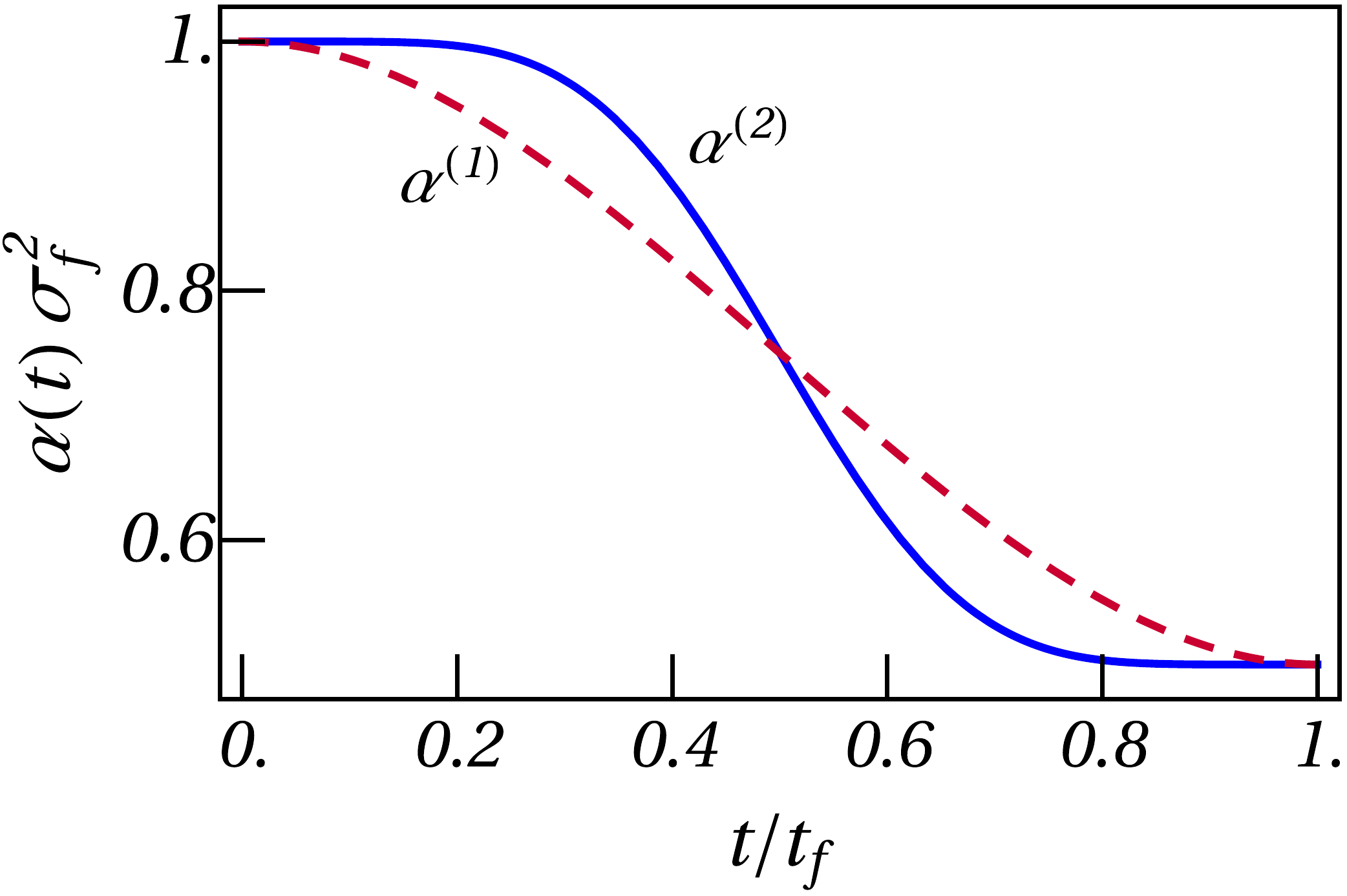}
\caption{Two different choices for $\alpha(t)$: $\alpha^{(1)}$ (in red) and $\alpha^{(2)}$ (in blue) as a function of rescaled time $t/t_{\mathrm{f}}$. The difference in their 
behaviour for small $t \simeq 0$ and for $t \simeq t_{\mathrm{f}}$
is at the root of their distinct efficiency.
}
\label{fig:alpha}
\end{figure}

Before putting the scheme to the test for a one dimensional confined system $-L/2 \leq x \leq L/2$,
we discuss the choice of the inverse variance of the colloid's position through $\alpha(t)$, with which the whole procedure starts.
Any function that verifies the initial and final conditions of a prescribed $\alpha$ with vanishing first derivative, can be taken. The simplest choice is polynomial in time and 
reads
\begin{equation}
\alpha^{(1)}(t) = \frac{1}{2 k_{\mathrm{B}} T} \left(\frac{}{}\kappa_{\mathrm{i}} + (\kappa_{\mathrm{f}}-\kappa_{\mathrm{i}}) (3s^2-2s^3)\right),
\end{equation}
where $s=t/t_{\mathrm{f}}$. 
Here again, other choices that fulfill the 
boundary conditions are possible and
a more suitable variant turns out to be
\begin{eqnarray}
\label{eq:alpha2}
 &&\alpha^{(2)}(t) = \frac{1}{2 k_{\mathrm{B}} T} \left(\frac{}{}\kappa_{\mathrm{i}} + 
 (\kappa_{\mathrm{f}}-\kappa_{\mathrm{i}})(1716 s^7 - 9009 s^8 + 20020 s^9 \right.\nonumber\\
 && \left. - 24024 s^{10} + 16380 s^{11} - 6006 s^{12} + 924 s^{13})\frac{}{}\right).
\end{eqnarray}
Indeed, the protocol stemming from $\alpha^{(1)}$ is less accurate than its $\alpha^{(2)}$-counterpart:
the reason for this lies in the smoothness of $\alpha^{(2)}(t)$, rather than $\alpha^{(1)}(t)$, at initial and final times, as illustrated in Fig.~\ref{fig:alpha}:
the first 6 derivatives of $\alpha^{(2)}$ do vanish at initial and final times, as compared to the first derivative only with $\alpha^{(1)}$. This smoothness leads to a soft running of concentration before $t_{\mathrm{f}}$. Then, the ions are offered more time to reach the uniform state before finishing the process, reducing the ion diffusion effect on the colloid's final state.
This results in an enhanced stability of the whole scheme, as explicit comparison for both protocols reveals, see~\ref{appen:res_alpha1} for the $\alpha^{(1)}$
calculation.

Besides, it proves convenient to take
\begin{equation}
\kappa_{\mathrm{op}}(t) \,=\, 2 k_{\mathrm{B}}T \alpha(t),
\label{eq:2alpha}
\end{equation}
so that from Eq. \eqref{eq:kappa} we get 
\begin{equation}
    \kappa_{\mathrm{DP}} \,=\, \frac{\dot{\alpha}(t)}{2 \alpha(t)} \gamma 
    \label{eq:kapDP}
\end{equation}
Note that this choice, in the quasi-static limit where
$t_{\mathrm{f}}\to \infty$, would lead to a consistent 
adiabatic protocol without the need of any diffusiophoretic 
assistance~\cite{comment101}. In doing so, it can be considered that the 
non adiabatic contribution to the protocol is entirely borne 
by the diffusiophoretic drive.
Then, from Eq.~\eqref{eq:C_0}, $C(0,t)$ takes the simple form
\begin{equation}
C(0,t) = C_0 \left(\frac{\alpha}{\alpha_{\mathrm{i}}}\right)^
{-\frac{n D_{\mathrm{s}}}{2\Gamma_{\mathrm{c}}}}.
\label{eq:x2_C_0}
\end{equation}
Equation \eqref{eq:kappa} subsequently indicates how the total stiffness $\kappa$ acting on the colloid should depend on time. 
This stiffness has an optical and a diffusiophoretic contribution. The optical contribution has been chosen to be $\kappa_{\mathrm{op}}=2k_{\mathrm{B}}T\alpha(t)$,
and it remains to be seen what is the corresponding 
time dependence for $\kappa_{\mathrm{DP}}$. This is 
a rather subtle task, since what is supposed
to be experimentally controllable is the salt 
concentration $\phi(t)$ at the boundary of the domain (assumed for simplicity to be position independent
while in principle, it could also exhibit 
a non trivial spatial dependence). Thus, we look for the $\phi(t)$ that would guarantee that 
$U_{\mathrm{DP}}$ in Eq.~\eqref{eq:UDF} is quadratic close
to the origin, and with curvature $\kappa_{\mathrm{DP}}(t)$.

\begin{figure}[!ht]
\centering  \includegraphics[width=1\linewidth]{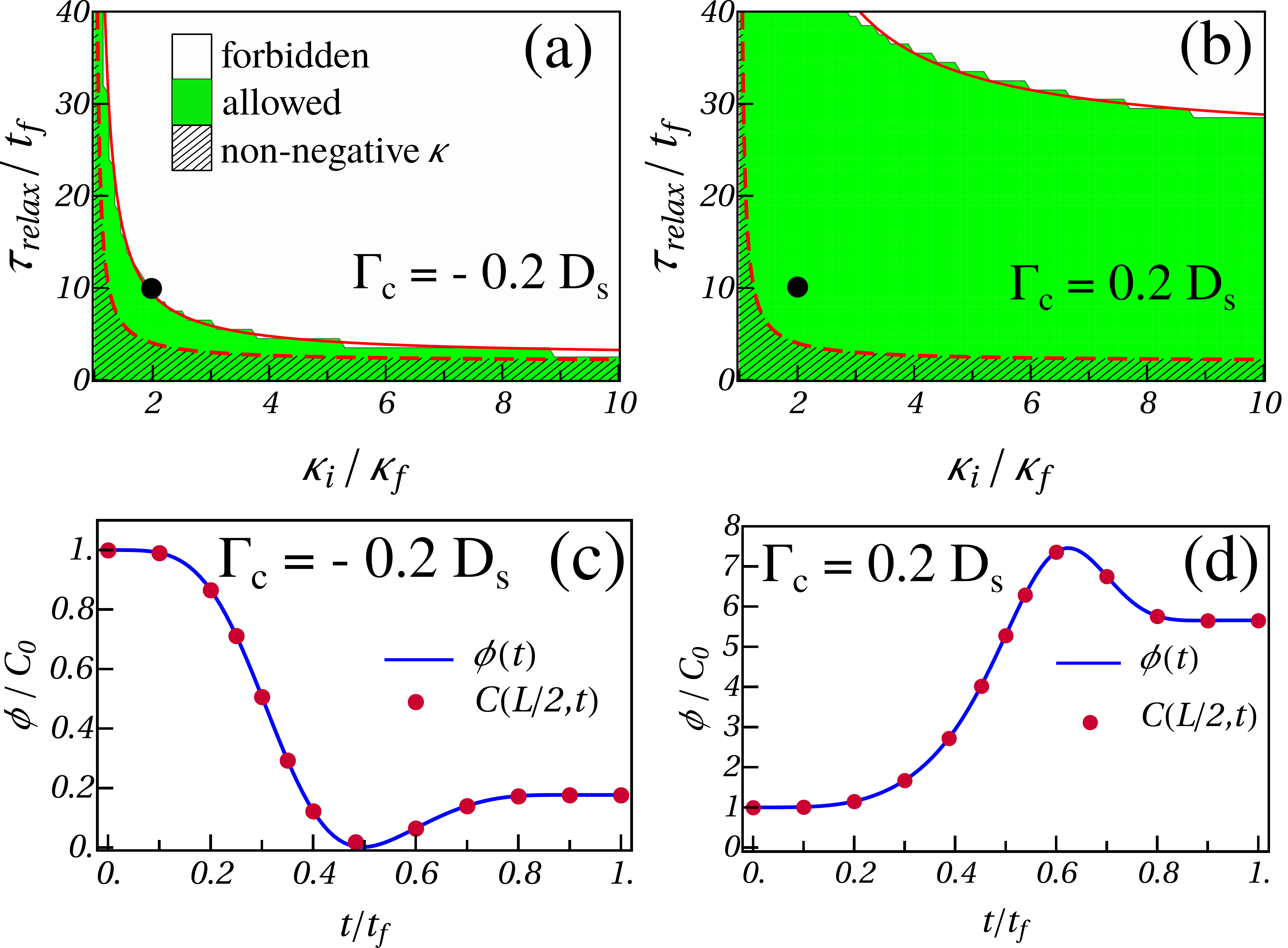}
\caption{\label{fig:phase_a2_1D_x2}
(a -- b) Phase plots of accessible state points as a function of ${\cal A}=\tau_{\mathrm{relax}}/t_{\mathrm{f}}$ and the 
compression factor $\kappa_{\mathrm{i}}/\kappa_{\mathrm{f}}$, for different phoretic mobilities:
(a) $\Gamma_{\mathrm{c}} = - 0.2~D_{\mathrm{s}}$ and (b) $\Gamma_{\mathrm{c}} = 0.2~D_{\mathrm{s}}$. 
The  green regions show the values for which $\phi(t)$ is positive at all times, i.e. in the whole interval $[0,t_{\mathrm{f}}]$. Conversely,
the white regions are  for those states for which $\phi(t)$ takes a negative value at least once during the protocol.
The plots correspond to $\alpha(t) = \alpha^{(2)}(t)$ 
given by Eq.~\eqref{eq:alpha2} and optical stiffness $\kappa_{\mathrm{op}} = 2 k_{\mathrm{B}}T \alpha(t)$,
with $D_{\mathrm{c}} = 0.002~D_{\mathrm{s}}$, and a system size $L=10~\sigma_{\mathrm{f}}$.
The hatched region shows the accessible states obtained by any ESE protocol
with an ever non--negative stiffness (given by Eq. (34) in Ref.~\cite{plata2019optimal}). The 
non-hatched green region therefore 
helps visualizing the benefits of the present protocols
over that of Ref. \cite{plata2019optimal}.
Panels
(c -- d) show the boundary salt concentration $\phi(t)$ for the same parameters, for ${\cal A} = 10$ and $\kappa_{\mathrm{i}}/\kappa_{\mathrm{f}} = 2$, corresponding to the black dots in panels (a-b). This quantity 
is supposed to be externally controlled in the experiment; it 
is our driving field.
The blue curves show the results obtained by the Taylor expansion \eqref{eq:phi_Taylor}, truncated at order $j=16$:
it is the computed driving concentration to be applied to the system's boundary. On the other hand, the 
red dots show the concentration obtained by numerically solving the diffusion equation~\eqref{eq:diffusion}, i.e., $C(\pm L/2, t)$, imposing $\phi(t)$ at the boundary.
The two functions should coincide, as they do here; a lack 
of accuracy or self--consistency
in the treatment would lead to a mismatch.
}
\end{figure}

Before calculating $\phi(t)$, we discuss how fast the procedure can be run.
As pointed out in section~\ref{sec:ESE_DP_G}, enforcing the rule given in Eq.~\eqref{eq:kappa} may give rise to negative concentrations, particularly so when high speed is sought:
this sets a lower bound for the operating time $t_{\mathrm{f}}$ 
when a given compression ratio $\kappa_{\mathrm{i}}/\kappa_{\mathrm{f}}$ is 
targeted or conversely, an upper bound for $\kappa_{\mathrm{i}}/\kappa_{\mathrm{f}}$
when $t_{\mathrm{f}}$ is fixed.
These bounds are shown in Fig.~\ref{fig:phase_a2_1D_x2}(a) and (b).
The green regions show the accessible final states where $\phi(t)$ is positive during the whole process $0\leq t \leq t_{\mathrm{f}}$. However, the states in the white regions are not achievable since they lead to a negative $\phi(t)$  at least once in the time interval $0\leq t \leq t_{\mathrm{f}}$.

Figure~\ref{fig:phase_a2_1D_x2} illustrates that the extent 
of the allowed region is smaller for negative phoretic
mobility than for positive ones. This is related to the sketch of Fig.~\ref{fig:con_model}, and the fact that when $\Gamma_{\mathrm{c}}<0$, the smaller salt concentrations should be
in the vicinity of the domain's boundary (inwards salt gradient), see also panels (c) and (d) in Fig.~\ref{fig:phase_a2_1D_x2}. The state point for panel (c) lies 
on the boundary of the allowed region in panel (a) (see the black dot in panel (a)) and it is indeed observed that the 
time profile $\phi(t)$ in panel (c) does vanish 
for $t\simeq t_{\mathrm{f}}/2$. The gist of our proposal 
is to compute the driving $\phi(t)$ to be imposed 
as the salt concentration at the boundary. Fig \ref{fig:phase_a2_1D_x2}-(c) and (d) provide us with an explicit answer for the illustrative parameters chosen.
When $C(0,t)$ is known, we use the truncated expansion 
\eqref{eq:phi_Taylor} to get the boundary concentration
$\phi(t)$; truncation at order $j=16$ leads to an acceptable balance between accuracy and simplicity.


In Fig.~\ref{fig:phase_a2_1D_x2}, the separatrix between the green (allowed) and white (forbidden) regions 
determines a ``speed limit'' of the expansion~\cite{plata2020finite}. This separatrix is the locus of state points for which the required salt concentration 
$\phi(t)$ becomes negative.
 The fact that $\phi(t_0)=0$ for some time $t_0$ 
(with $0 \leq t_0 \leq t_{\mathrm{f}}$) gives a relation between ${\cal A}$ and $\kappa_{\mathrm{f}}$,
of the form
\begin{equation}
{\cal A} \simeq \frac{C_1}{\kappa_{\mathrm{f}} - \kappa_{\mathrm{i}}} + C_2,
\end{equation}
where the constants $C_1$ and $C_2$ depend on the choice of $\alpha$ and the phoretic mobility. This yields the boundary line between the allowed and forbidden regions in Fig.
\ref{fig:phase_a2_1D_x2}.

\subsection{Checking consistency}

To assess the protocol quality, we need to check for the results consistency. Indeed, our calculation 
of the driving field $\phi(t)$ relies on an expansion
that connects information pertaining to the system center
to information at the boundary.
We thus solve numerically the diffusion equation for salt density, and obtain the concentration profile $C(x,t)$,
from which all quantities of interest follow, from colloidal forces to observables like the colloidal position variance $\sigma^2$.
A first stringent test is to verify that the target
$\phi(t)$, worked out analytically, coincides with the 
salt concentration at the boundary, measured from the 
numerical solution $C(x,t)$. This self-consistency 
requirement is well obeyed in  Fig.~\ref{fig:phase_a2_1D_x2}(c) and (d),
see the blue curves and red dots.
This confirms that the proposed $\alpha(t)$ is a suitable choice, and that we have retained enough 
terms in the Taylor truncation.~\ref{appen:res_alpha1} indicates that self-consistency is somewhat better when working with $\alpha^{(2)}$
rather than $\alpha^{(1)}$.

\begin{figure}[!ht]
\centering \includegraphics[width=0.98\linewidth]{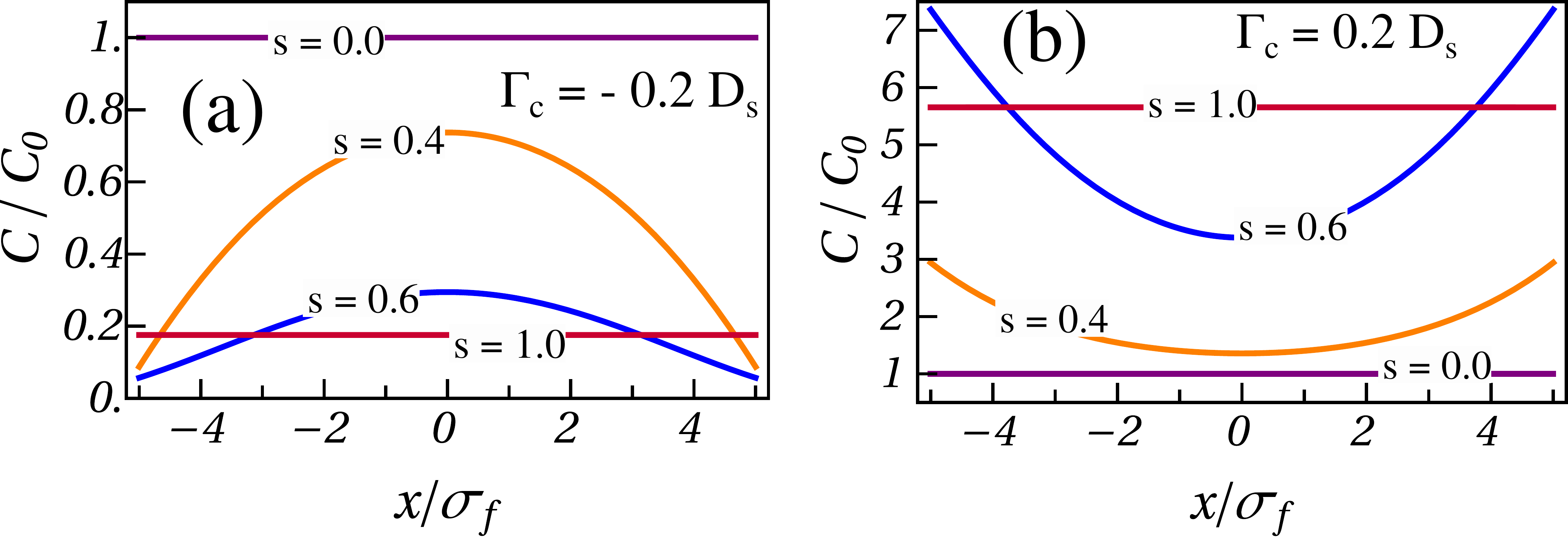}
\caption{\label{fig:conc_1D_x2} 
Salt concentration $C(x,t)$ at different scaled times $s=t/t_{\mathrm{f}}$ for two phoretic mobilities (a) $\Gamma_{\mathrm{c}} = - 0.2 D_{\mathrm{s}}$ (b) $\Gamma_{\mathrm{c}} = 0.2 D_{\mathrm{s}}$ obtained by solving the diffusion equation~\eqref{eq:diffusion}
numerically, imposing the boundary condition given by $\phi(t)$ in Fig~\ref{fig:phase_a2_1D_x2}(c--d).
}
\end{figure}

The numerically computed ions concentration profiles $C(x,t)$ are shown in Fig~\ref{fig:conc_1D_x2}.
The concentration is uniform at the initial and final times.
In between, a gradient sets in, which is towards the center for $\Gamma_{\mathrm{c}}<0$ 
and away from the center for $\Gamma_{\mathrm{c}}>0$, as sketched in Fig.~\ref{fig:con_model}.

\begin{figure}[!ht]
\centering \includegraphics[width=0.75\linewidth]{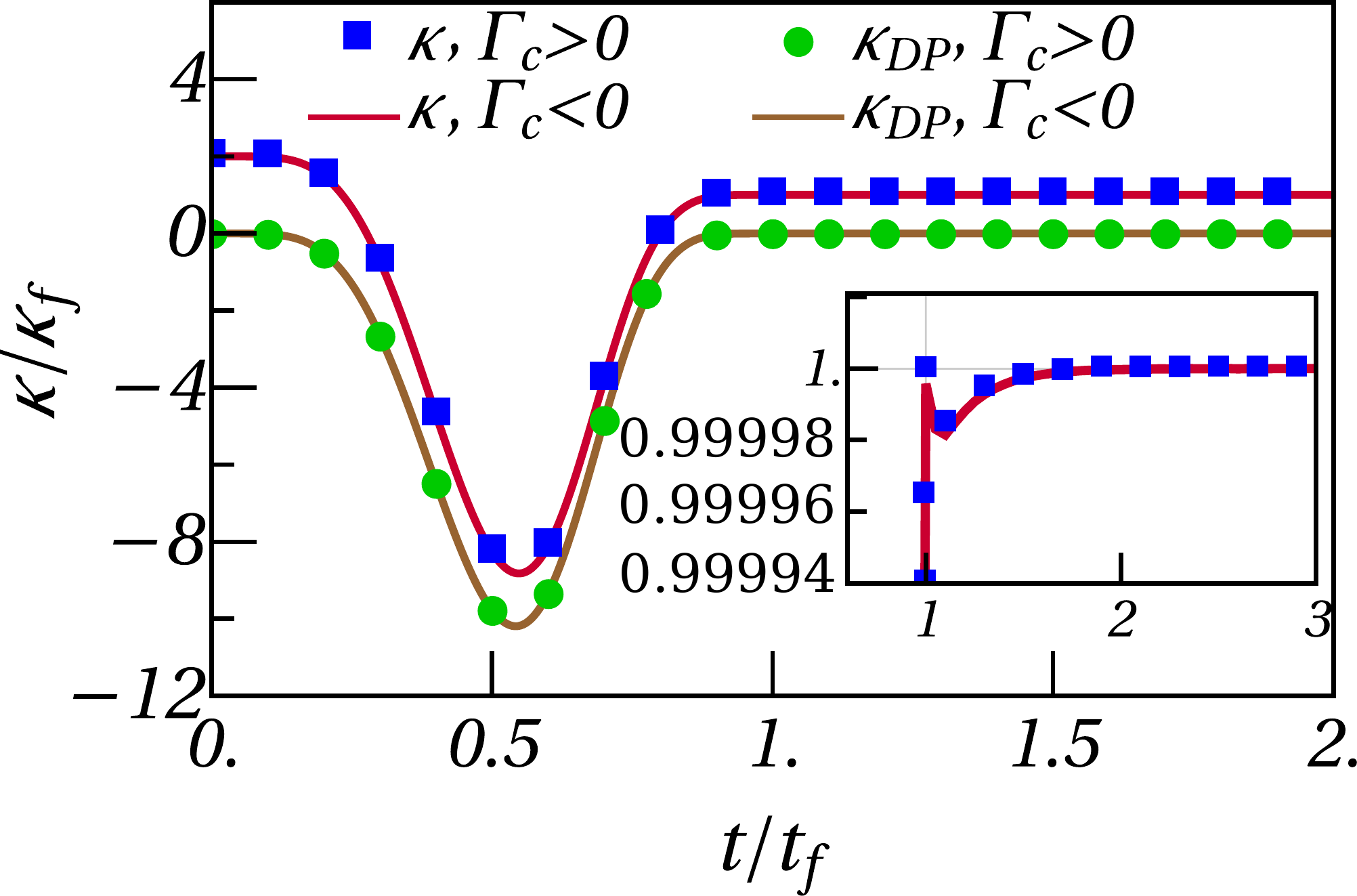}
\caption{\label{fig:kappa_all_1D_x2} 
Total ($\kappa$) and diffusiophoretic ($\kappa_{\mathrm{DP}}$) stiffnesses
calculated by solving the diffusion equation \eqref{eq:diffusion} numerically, as a function of time for 
the same parameters as Fig.~\ref{fig:phase_a2_1D_x2},
i.e. in particular $\Gamma_{\mathrm{c}} = \pm 0.2~D_{\mathrm{s}}$.
The inset shows the total stiffness at late times. The deviation of $\kappa$ from the target value after finishing the process is negligible.
}
\end{figure}

The next results that should be bench-marked are the stiffnesses. 
The total stiffness $\kappa(t)$ and diffusiophoretic stiffness $\kappa_{\mathrm{DP}}$ are shown in Fig.~\ref{fig:kappa_all_1D_x2}.
Starting from $\kappa_{\mathrm{i}}=2\kappa_{\mathrm{f}}$, we see that $\kappa(t)$ reaches its target value $\kappa_{\mathrm{f}}$ at $t=t_{\mathrm{f}}$. 
It is observed that $\kappa_{\mathrm{DP}}(t)$, which originates from the salt concentration gradient, vanishes at the initial and final times because of the uniform salt distribution. In between, it is negative as embodied in Eq.~\eqref{eq:kappa_DP}.
Besides, by construction, we have chosen that not only $\kappa$ but also 
$\kappa_{\mathrm{op}}$ be directly related to 
$\alpha(t)$. Hence, the dynamics of $\kappa$ and
$\kappa_{\mathrm{op}}$ are independent of the sign of the
mobility $\Gamma_{\mathrm{c}}$, as illustrated in Fig.~\ref{fig:kappa_all_1D_x2}. Note from Fig.~\ref{fig:conc_1D_x2} that the required salt profiles, however,
do strongly depend on the sign of $\Gamma_{\mathrm{c}}$.
Furthermore, a more careful inspection with the inset plots
reveals that at $t=t_{\mathrm{f}}$, the salt profile
is not completely steady, which results in an undesired force
on the colloids: as discussed in section~\ref{sec:ESE_DP_G}, the diffusion of ions for $t>t_{\mathrm{f}}$ leads to 
a small error for the measured stiffness, compared to
the target value. 
A perfect protocol would lead to $\kappa=\kappa_{\mathrm{f}}$
exactly at $t_{\mathrm{f}}$, without any variation afterwards. Note however the smallness of the mismatch (distance to unity) in the inset of Fig.~\ref{fig:kappa_all_1D_x2}.

\begin{figure}[!ht]
\centering  \includegraphics[width=0.65\linewidth]{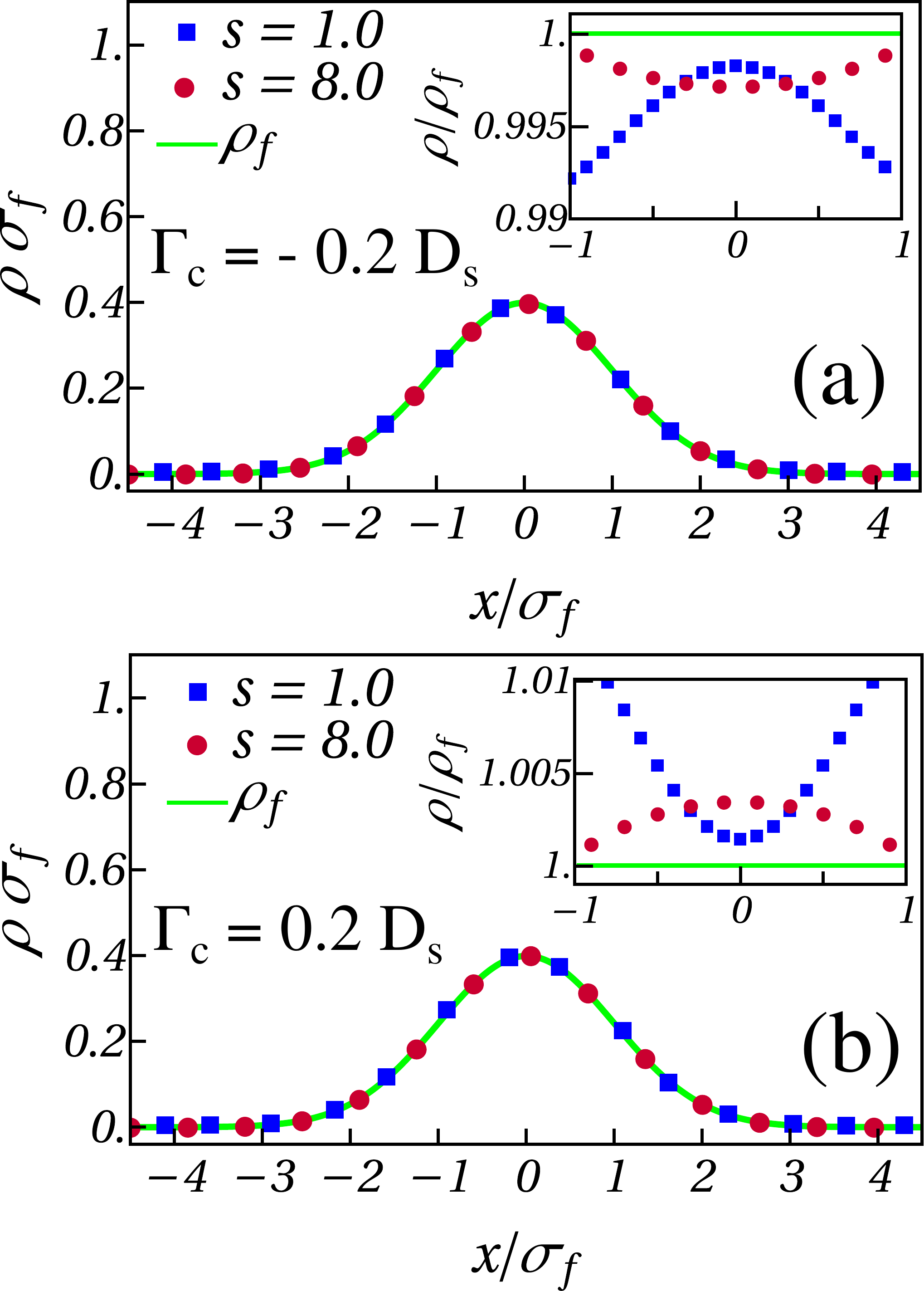}
\caption{\label{fig:pdf_1D_x2} 
Colloidal density $\rho_{\mathrm{num}}(x,t)$ calculated by solving the Fokker-Planck equation~\eqref{eq:fokker} numerically, at the final time $s = t/t_{\mathrm{f}} =1.0$ and some time after finishing the process ($s = 8.0$). 
The driving potential acting on the colloids, $U$ in Eq.~\eqref{eq:U_x2}, is obtained from solving numerically the salt diffusion equation \eqref{eq:diffusion}.
The achieved density overlaps with the target Gaussian distribution  
$\rho_{\mathrm{f}} = \rho(x,t_{\mathrm{f}})$ given by Eq.~\eqref{eq:G_pdf}. At other times, the agreement is equally good. 
The panels correspond to (a) $\Gamma_{\mathrm{c}} = -0.2~D_{\mathrm{s}}$ and (b) $\Gamma_{\mathrm{c}} = 0.2~D_{\mathrm{s}}$ for 
$\alpha(t)$ and $\kappa_{\mathrm{op}}(t)$ given by Eq.~\eqref{eq:alpha2} and 
Eq.~\eqref{eq:2alpha}, respectively.
The insets represent  $\rho_{\mathrm{num}}/\rho_{\mathrm{f}}$ near the system center. The densities differ slightly from the target Gaussian distribution.
}
\end{figure}

Having the concentration profile $C(x,t)$, one can solve numerically the Fokker-Planck equation, Eq.~\eqref{eq:fokker},
to obtain the colloid density $\rho_{\mathrm{num}}(x,t)$
in order to compare with the target Gaussian distribution $\rho(x,t)$ given by Eq.~\eqref{eq:G_pdf}. 
The results for the density in different times are demonstrated in Fig.~\ref{fig:pdf_1D_x2}.
The density keeps its Gaussian distribution during the whole process. One 
observes that at the time $t_{\mathrm{f}}$, and also after that $t>t_{\mathrm{f}}$, the density differs slightly
from the final target distribution. However, as the inset figures show, the differences are small.

\begin{figure}[!ht]
\centering  \includegraphics[width=0.65\linewidth]{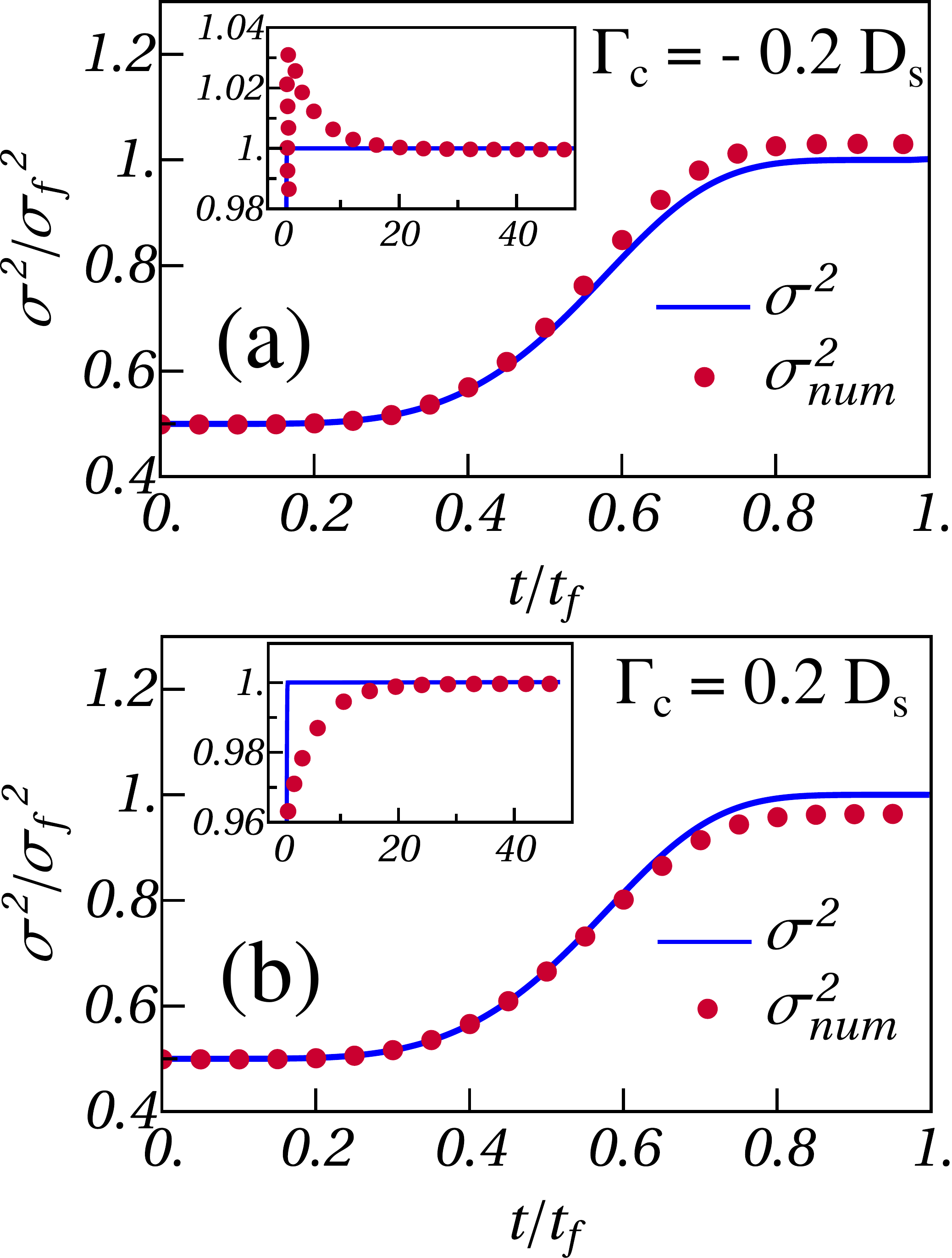}
\caption{\label{fig:var_1D_x2}
Variance $\sigma^2$ of the colloidal position distribution as a function of time.
(a) $\Gamma_{\mathrm{c}} = -0.2~D_{\mathrm{s}}$ and (b) $\Gamma_{\mathrm{c}} = 0.2~D_{\mathrm{s}}$ for 
$\alpha(t)=\alpha^{(2)}(t)$ and $\kappa_{\mathrm{op}}(t)$ given by Eq.~\eqref{eq:alpha2} and 
Eq.~\eqref{eq:2alpha}.
Same parameters as in Fig.~\ref{fig:phase_a2_1D_x2}.
The variance calculated numerically, i.e., $\sigma_{\mathrm{num}}^2 = \int x^2 \rho_{\mathrm{num}}(x,t)~dx$, is compared to the target $1/[2\alpha(t)]$; it deviates slightly from this quantity, with a maximum mismatch at $t_{\mathrm{f}}$.
The insets show the evolution for longer times, where a stable value is reached for $t \simeq 20 \, t_{\mathrm{f}}$, which shows a good match with the target value.
%
%
}
\end{figure}

Finally, we compute the variance of the colloid $\sigma^2(t)$ to ensure that the target expansion is realized.
Recall that $\sigma^2 = (2\alpha)^{-1}$ and that 
the time dependence of $\alpha$ is chosen from the outset.
The variance is plotted as a function of time for two phoretic mobilities $\Gamma_{\mathrm{c}} = \pm 0.2~D_{\mathrm{s}}$
in Fig.~\ref{fig:var_1D_x2}. 
The variance calculated numerically, i.e. $\int_{-\infty}^{+\infty} x^2 \rho_{\mathrm{num}}(x,t)~dx$, 
is close to the target one $1/(2\alpha(t))$; however, it deviates slightly
at long times. This is due to the density mismatch 
displayed in  Fig.~\ref{fig:pdf_1D_x2}. 
It turns out that $\sigma^2$ is a rather sensitive
probe for the small imperfections of the protocol,
and reveals possible defects more clearly than
the quantities reported hitherto.
At any rate, the difference between the desired and
observed variances is less than a few percents at all time,
and stabilizes to a value very close to the target variance.
Besides, although we have assumed that the colloid density is Gaussian, the non-quadratic terms in the potential, Eq.~\eqref{eq:UDF},  may cause non-Gaussianity. In~\ref{appen:nonG_G}, we quantify the deviations from Gaussian distribution and show that they also remain relatively small.

\section{Manipulating non-Gaussian states\label{sec:ESE_DP_non_G_pdf}}
The previous sections were devoted to shortcutting the dynamics of a system confined in a harmonic potential, enforcing the colloidal state to be Gaussian at all times.  
We address here the situation of a non-harmonic confinement.
A non-Gaussian distribution ensues.
We will consider a quartic confining potential $U_{\mathrm{eq}} \sim x^4$ as an illustrative example.
At equilibrium, we then have  $\rho_{\mathrm{eq}} \sim \exp(-\alpha x^4)$.
The aim is to drive the system from an initial state with variance 
$\sigma_{\mathrm{i}}$ to a final state with increased variance $\sigma_{\mathrm{f}}$, in a processing time $t_{\mathrm{f}}$, arbitrarily shorter than the relaxation time $\tau_{\mathrm{relax}} = \sigma_{\mathrm{f}}^2/D_{\mathrm{c}}$.
To this end, we again follow the ESE procedure, as done in section~\ref{sec:ESE_DP_G} and in  Ref.~\cite{martinez2016engineered}, 
working out in one dimension,
and driving the system through the control of salt concentration $\phi(t)$ at its boundaries ($x=\pm L/2$).
We will demand that the colloid density be at all times of the form
\begin{equation}
\label{eq:x4_pdf}
\rho(x,t) = \frac{\alpha^{1/4}(t)}{2\Gamma(5/4)} \exp(-\alpha(t) x^4),
\end{equation}
where $\Gamma(x)$ represents the Gamma function and the prefactor of the exponential stems
from normalization.
Variance and kurtosis of the distribution are 
\begin{equation}
\sigma^2 =  \frac{N}{2\sqrt{\alpha}}, \qquad
{\cal K} = \frac{\langle (x-\langle x\rangle)^4 \rangle}{\sigma^4}-3 
\simeq 0.812
\end{equation}
where $N = \Gamma(3/4)/ (2\Gamma(5/4)) \simeq 0.678$ is a numerical factor.
It can be shown that a suitable driving potential reads
\begin{equation}
\label{eq:x4poten}
U(x,t) = \frac{1}{4} \delta(t) x^4 + \frac{1}{2} \kappa(t) x^2.
\end{equation}
This potential should match the equilibrium potential in $x^4$ at the initial and final times, so that $\kappa(0)=\kappa(t_{\mathrm{f}}) = 0$.
The function $\alpha(t)$ can be chosen arbitrarily, fulfilling the equilibrium boundary conditions
$\alpha(0)= \delta_{\mathrm{i}}/(4 k_{\mathrm{B}}T)$ and $\alpha(t_{\mathrm{f}})= \delta_{\mathrm{f}} / (4 k_{\mathrm{B}}T)$. Thus, the variance of final state is $\sigma_{\mathrm{f}}^2 = N \sqrt{k_{\mathrm{B}}T/\delta_{\mathrm{f}}} \equiv N \sigma_{\mathrm{0}}^2$. Moreover. we choose $\alpha(t)$ such that $\dot{\alpha}(0) =\dot{\alpha}(t_{\mathrm{f}})=0$ to have a smooth function at initial and final times. As with the harmonic situation, we restrict to continuous functions.

Substituting $\rho(x,t)$ and $U(x,t)$ into the Fokker-Planck equation \eqref{eq:fokker} gives:
\begin{eqnarray}
\frac{\dot{\alpha}}{4\alpha} - \dot{\alpha} x^4 &&=  D_{\mathrm{c}} (-12 \alpha x^2 + 16 \alpha^2 x^6) \nonumber\\
&&+ \frac{1}{\gamma} (3 \delta x^2 + \kappa) - \frac{4 \alpha}{\gamma} x^3 (\delta x^3 + \kappa x),
\end{eqnarray}
which is valid for any position $x$. This implies that
\begin{equation}
\label{eq:x4_del_ka}
 \kappa(t) = \frac{\gamma}{4} \frac{\dot{\alpha}(t)}{\alpha(t)}, \qquad \delta(t) = 4 k_{\mathrm{B}} T \alpha(t).
\end{equation}
Therefore, choosing an appropriate function for $\alpha(t)$ also sets the stiffness coefficients $\kappa(t)$ and $\delta(t)$, through Eq.~\eqref{eq:x4_del_ka}.
A decompression results in having 
$\dot{\alpha} < 0$, so that $\kappa(t) < 0$. As a result, the potential $U$ 
in \eqref{eq:x4poten} features a double-well form
for $0<t<t_{\mathrm{f}}$~\cite{martinez2016engineered} ($\alpha$ and thus $\delta$ are positive).

The density of bathing salt solution $C(x,t)$ satisfies the diffusion equation \eqref{eq:diffusion} with boundary condition given by Eq.~\eqref{eq:IBC}. Here again, the goal is 
to work out the salt concentration at the boundary of the system, that will lead to the diffusiophoretic force in the
bulk that is precisely of the form leading to the desired
dynamics in Eq.~\eqref{eq:x4_pdf}. As before, the colloidal dynamics is enslaved to that of salt through the Fokker-Planck equation, while salt itself obeys pure diffusion. 

With both optical and diffusiophoretic potentials, the total potential acting on the colloid is:
\begin{equation}
U = U_{\mathrm{op}} + U_{\mathrm{DP}} = \frac{1}{2} \kappa_{\mathrm{op}} x^2 + \frac{1}{4}\delta_{\mathrm{op}} x^4 - \gamma \Gamma_{\mathrm{c}} \ln C(x,t).
\end{equation}
We thus consider here that the non linearity of the optical
trapping leads to a quartic term in $U$, in addition to the
standard harmonic term. While this is routine experimental 
practice for the quadratic term in $\kappa_{\mathrm{op}}$, we also 
assume that the quartic term can be tuned over time by the operator,
which may prove difficult.
The present analysis, compared to that of the previous sections,
is for this reason more speculative.
In line with the previous analysis, we 
take $\kappa_{\mathrm{op}}(t)$ and $\delta_{\mathrm{op}}(t)$ as non-negative for all times.
In addition, we assume isotropy and analyticity, which implies here 
$\partial^j_{x} \ln C|_{x=0} =0$ for $j$ odd.
Thus, the expansion near the system center gives
\begin{eqnarray}
U &&\simeq \frac{1}{2}\left(\frac{ }{} \kappa_{\mathrm{op}} - \gamma \Gamma_{\mathrm{c}}  \partial^2_{x}\ln C(x,t)|_{x=0} \right) x^2\nonumber\\
&&+ \frac{1}{4} \left(\frac{ }{}\delta_{\mathrm{op}} - \frac{1}{6} \gamma \Gamma_{\mathrm{c}} \partial^4_{x}\ln C(x,t)|_{x=0} \right) x^4.
\end{eqnarray}
Then, the diffusiophoretic stiffnesses are
\begin{eqnarray}
\label{eq:x4_kappa_DP}
\kappa_{\mathrm{DP}}(t) &&\equiv \kappa(t) -\kappa_{\mathrm{op}}(t) = - \gamma \Gamma_{\mathrm{c}}  \partial^2_{x}\ln C(x,t)|_{x=0},\nonumber\\
\delta_{\mathrm{DP}}(t) &&\equiv \delta(t) -\delta_{\mathrm{op}}(t) = - \frac{1}{6} \gamma \Gamma_{\mathrm{c}} \partial^4_{x}\ln C(x,t)|_{x=0}.
\end{eqnarray}
Combining with Eq.~\eqref{eq:x4_del_ka} and using the diffusion equation \eqref{eq:diffusion}, we obtain the following differential equation 
for the density at the center $C(0,t)$ and optical stiffnesses:
\begin{eqnarray}
\label{eq:x4_delta_DP}
&&\kappa_{\mathrm{DP}}(t) = 
-\frac{\gamma \Gamma_{\mathrm{c}}}{{D_\mathrm{s}}}
\partial_{t}\ln C(0,t)
= -\kappa_{\mathrm{op}}(t) + \frac{\gamma}{4}
\frac{\dot{\alpha}(t)}{\alpha(t)}\nonumber\\
&&\delta_{\mathrm{DP}}(t) = 
-\frac{1}{6} \frac{\gamma \Gamma_{\mathrm{c}}}{D_\mathrm{s}^2}
\left(-2 (\partial_{t}\ln C(0,t))^2 \frac{}{ } + \partial^{2}_{t}\ln C(0,t)\right) = \nonumber\\
&&-\delta_{\mathrm{op}}(t) + 4 k_{\mathrm{B}}T \alpha(t).
\end{eqnarray} 
For a chosen $\alpha(t)$, the total quadratic and quartic stiffnesses $\kappa$ and $\delta$ are known, while 
$\kappa_{\mathrm{op}}$ or $\delta_{\mathrm{op}}$ are arbitrary function with no other constraint than being positive. 

\textit{Obtaining $\phi(t)$:}
As in the Gaussian case, there is some flexibility in the 
choice of $\alpha(t)$ and also $\kappa_{\mathrm{op}}(t)$. We take a similar function for $\alpha(t)$ as Eq.~\eqref{eq:alpha2},
that led previously to accurate protocols:
\begin{eqnarray}
\label{eq:alpha_nonG}
 &&\alpha^{(2)}(t) = \frac{1}{4 k_{\mathrm{B}} T} \left(\frac{}{}\delta_{\mathrm{i}} + 
 (\delta_{\mathrm{f}}-\delta_{\mathrm{i}})(1716 s^7 - 9009 s^8 + 20020 s^9 \right.\nonumber\\
 && \left. - 24024 s^{10} + 16380 s^{11} - 6006 s^{12} + 924 s^{13})\frac{}{}\right),
\end{eqnarray}
with again $s=t/t_{\mathrm{f}}$.
As far as $\kappa_{\mathrm{op}}(t)$ is concerned, it
proves convenient to choose
\begin{equation}
\label{eq:x4_kappaop}
\kappa_{\mathrm{op}}(t) = -\kappa(t) = - \frac{\gamma}{4} \frac{\dot{\alpha}(t)}{\alpha(t)}.
\end{equation}
Solving Eq.~\eqref{eq:x4_delta_DP} for $C(0,t)$ and $\delta_{\mathrm{op}}(t)$ yields
\begin{equation}
\label{eq:x4_C_0}
C(0,t) = C_0 \left(\frac{\alpha(t)}{\alpha_\mathrm{i}}\right)^{-\frac{D_{\mathrm{s}}}{2 \Gamma_{\mathrm{c}}}}
\end{equation}
and
\begin{equation}
\label{eq:x4_delta_op}
\delta_{\mathrm{op}}(t) = 4 k_{\mathrm{B}}T \alpha(t) + \frac{1}{12} \frac{k_{\mathrm{B}}T}{D_{\mathrm{c}} D_{\mathrm{s}}} \left(\frac{\dot{\alpha}^2}{\alpha^2} \left(1-\frac{D_{\mathrm{s}}}{\Gamma_{\mathrm{c}}}\right)-\frac{\ddot{\alpha}}{\alpha}\right).
\end{equation}
An interesting feature of the choice made for $\kappa_{\mathrm{op}}$
is that we recover the exact same dynamics for $C(0,t)$ as for 
Gaussian states, and thus we can recycle the corresponding results,
in particular the computation of $\phi(t)$. 
Indeed, our choice results in
\begin{equation}
    \kappa_{\mathrm{DP}} \,=\, \frac{\gamma}{2}\,\frac{\dot\alpha}{\alpha},
\end{equation}
which coincides with Eq. \eqref{eq:kapDP}.
As a consequence, the phase diagram in Fig.~\ref{fig:phase_a2_1D_x2}, together 
with the underlying salt density profile shown in Fig.~\ref{fig:conc_1D_x2}
also apply here, at the expense of replacing the compression factor $\kappa_{\mathrm{i}}/\kappa_{\mathrm{f}}$ by $\delta_{\mathrm{i}}/\delta_{\mathrm{f}}$.
Differences between 
the harmonic and non-harmonic confinement will nevertheless appear
when considering the force applied onto the colloids, as discussed next.

Note that we have assumed 
that $\delta_{\mathrm{op}}$ should remain positive. This is not guaranteed,
since the second term on the rhs of eq.~\eqref{eq:x4_delta_op}
may take negative values. Yet, remembering that $\alpha$ is of order $\sigma_{\mathrm{f}}^{-4}$, the ratio of the second and the first terms in eq.~\eqref{eq:x4_delta_op} is of order ${\cal A}^2 D_{\mathrm{c}}/(10 D_{\mathrm{s}})$. As in typical electrolytes $D_{\mathrm{c}}/D_{\mathrm{s}} \sim 10^{-3}$, 
we conclude that $\delta_{\mathrm{op}}>0$ as long as  
${\cal A} < 100$, or more generally 
${\cal A} < \sqrt{10 D_{\mathrm{s}}/D_{\mathrm{c}}}$. 
It should also be emphasized that as compared to the Gaussian situation, 
the present method is experimentally more challenging from the optical trapping
point of view, since not only the quadratic stiffness $\kappa_{\mathrm{op}}$ needs to be controlled in time,
but also the quartic stiffness $\delta_{\mathrm{op}}$.

\begin{figure}[!ht]
\centering \includegraphics[width=0.65\linewidth]{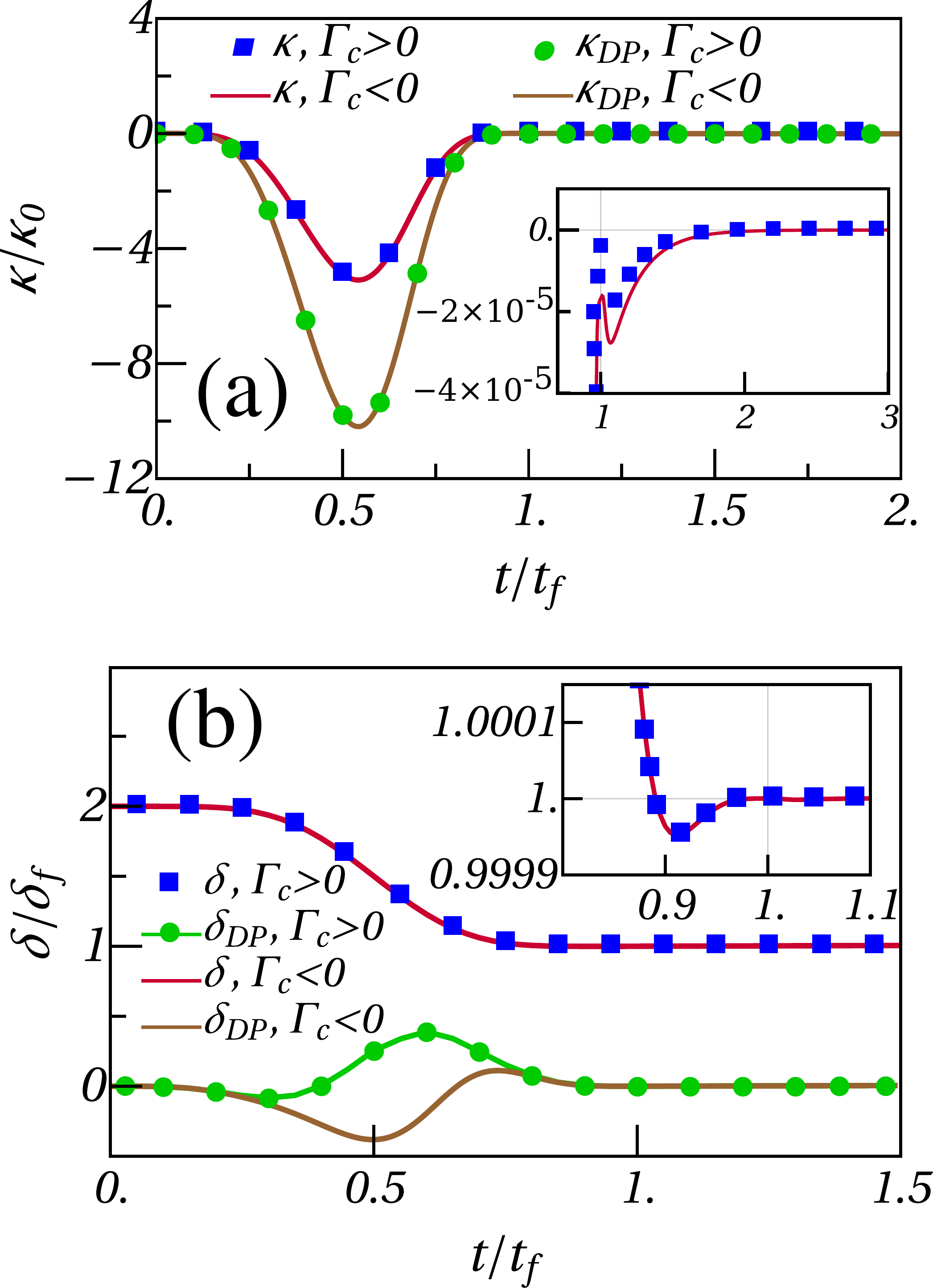}
\caption{\label{fig:kappa_all_1D_x4}
Non-Gaussian situation.
Total and diffusiophoretic stiffnesses, calculated by solving the diffusion equation \eqref{eq:diffusion} numerically,
as a function of rescaled time $t/t_{\mathrm{f}}$,
for the same parameters as in Fig.~\ref{fig:phase_a2_1D_x2}
and two phoretic mobilities 
$\Gamma_{\mathrm{c}} = \pm 0.2~D_{\mathrm{s}}$ where $\kappa_{\mathrm{0}} = N k_{\mathrm{B}} T/\sigma_{\mathrm{f}}^2$ and $\delta_{\mathrm{f}} = N^2 k_{\mathrm{B}} T/\sigma_{\mathrm{f}}^4$.
Here, $\alpha=\alpha^{(2)}$ as given in Eq.~\eqref{eq:alpha2}, from which $\kappa_{\mathrm{op}}(t)$, and  $\delta_{\mathrm{op}}$ follow, see Eqs.~\eqref{eq:x4_del_ka},\eqref{eq:x4_delta_DP}.
The insets show a close-up of the total stiffnesses near 
the final time. The deviation of $\kappa$ and  $\delta$ from the target value after finishing the process is negligible. 
}
\end{figure}

The stiffnesses $\kappa(t)$ and $\delta(t)$ are displayed in Fig.~\ref{fig:kappa_all_1D_x4}(a) and (b).
Once the diffusion equation has been solved for salt dynamics, 
the stiffnesses follow from Eqs. \eqref{eq:x4_del_ka} and \eqref{eq:x4_delta_DP}.
We can see that as required, not only $\kappa$ but also $\kappa_{\mathrm{DP}}$ (and thus $\kappa_{\mathrm{op}} = \kappa -\kappa_{\mathrm{DP}}$), do vanish both at $t=0$ and  $t=t_{\mathrm{f}}$. 
Besides, $\kappa_{\mathrm{DP}}$ and 
$\kappa$ are negative in the meantime, in order to generate the double-well
form for the potential, required to achieve the decompression. 
This is purely driven by diffusiophoresis.
The fact that $\kappa > \kappa_{\mathrm{DP}}$ signals that 
the optical contribution $\kappa_{\mathrm{op}}$ is always positive.
As with harmonic confinement, we have chosen the time dependence of 
$\kappa$ and $\kappa_{\mathrm{op}}$ to be blind to the sign 
of mobility $\Gamma_{\mathrm{c}}$. Thus, $\kappa_{\mathrm{DP}}$ inherits 
from this property. A similar comment holds for $\delta(t)$, while 
one needs to study $\delta_{\mathrm{DP}}$ to discriminate 
positive from negative $\Gamma_{\mathrm{c}}$.
The inset of Fig.~\ref{fig:kappa_all_1D_x4} indicates that a residual late time
dynamics slightly impinges on the forcing, and causes a small error in the measured  stiffnesses.

\begin{figure}[!ht]
\centering \includegraphics[width=0.65\linewidth]{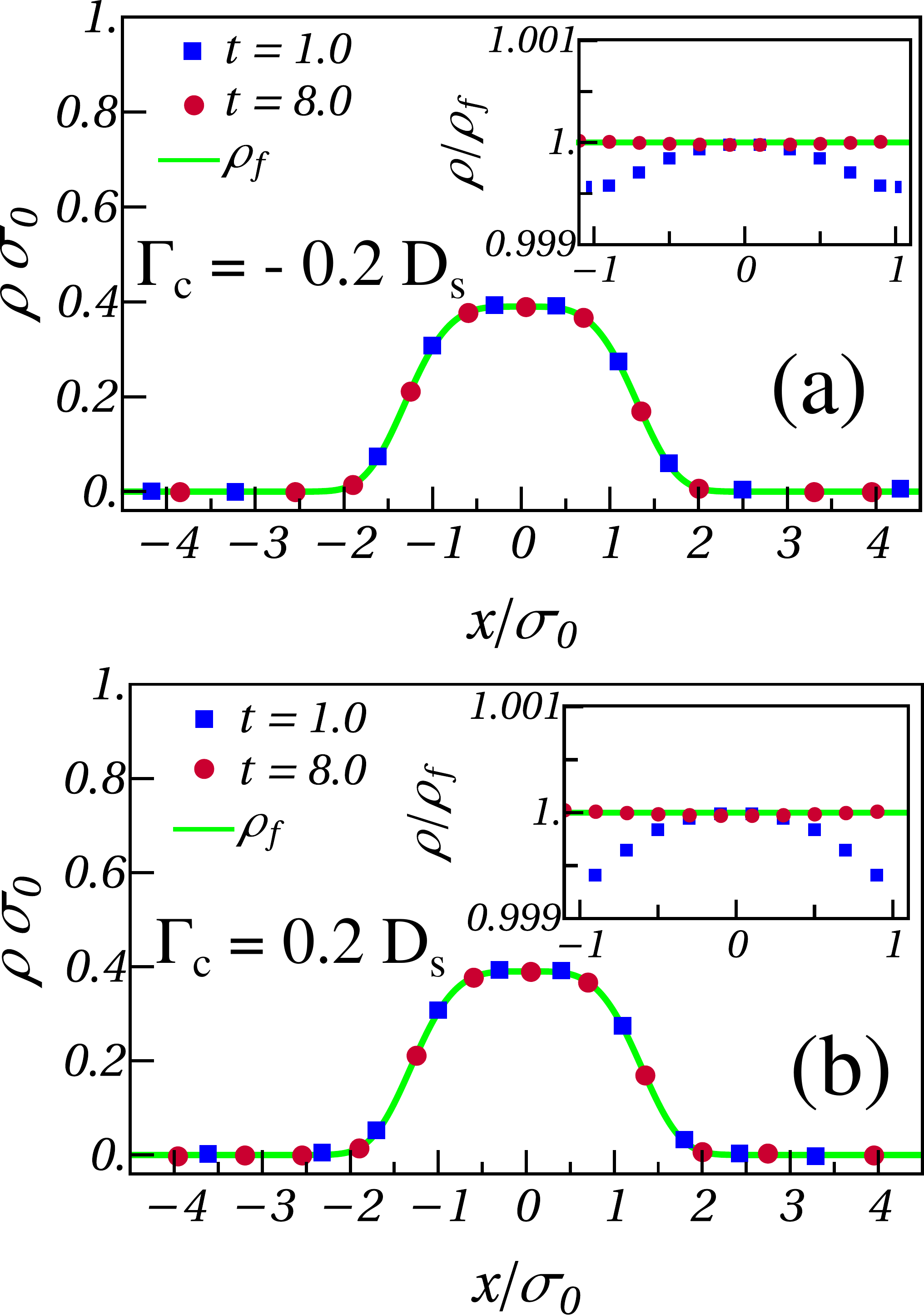}
\caption{\label{fig:pdf_1D_x4} 
Non-Gaussian situation.
Colloidal density $\rho_{\mathrm{num}}(x,t)$ calculated by solving the Fokker-Planck equation~\eqref{eq:fokker} numerically, for different times. 
$\rho_{\mathrm{f}}$ shows the target non-Gaussian distribution given by Eq.~\eqref{eq:x4_pdf}.
$\alpha^{(2)}(t)$, $\kappa_{\mathrm{op}}(t)$, and $\delta_{\mathrm{op}}(t)$ are given by Eqs.~\eqref{eq:alpha2},~\eqref{eq:x4_kappaop}, and~\eqref{eq:x4_delta_op}, respectively.
 The other parameters are the same as in Fig.~\ref{fig:phase_a2_1D_x2}.
The insets represent the ratio $\rho/\rho_{\mathrm{f}}$ with a zoom onto the center $x=0$. 
The densities very slightly differ from the target distribution.
}
\end{figure}

\begin{figure}[!ht]
\centering \includegraphics[width=0.65\linewidth]{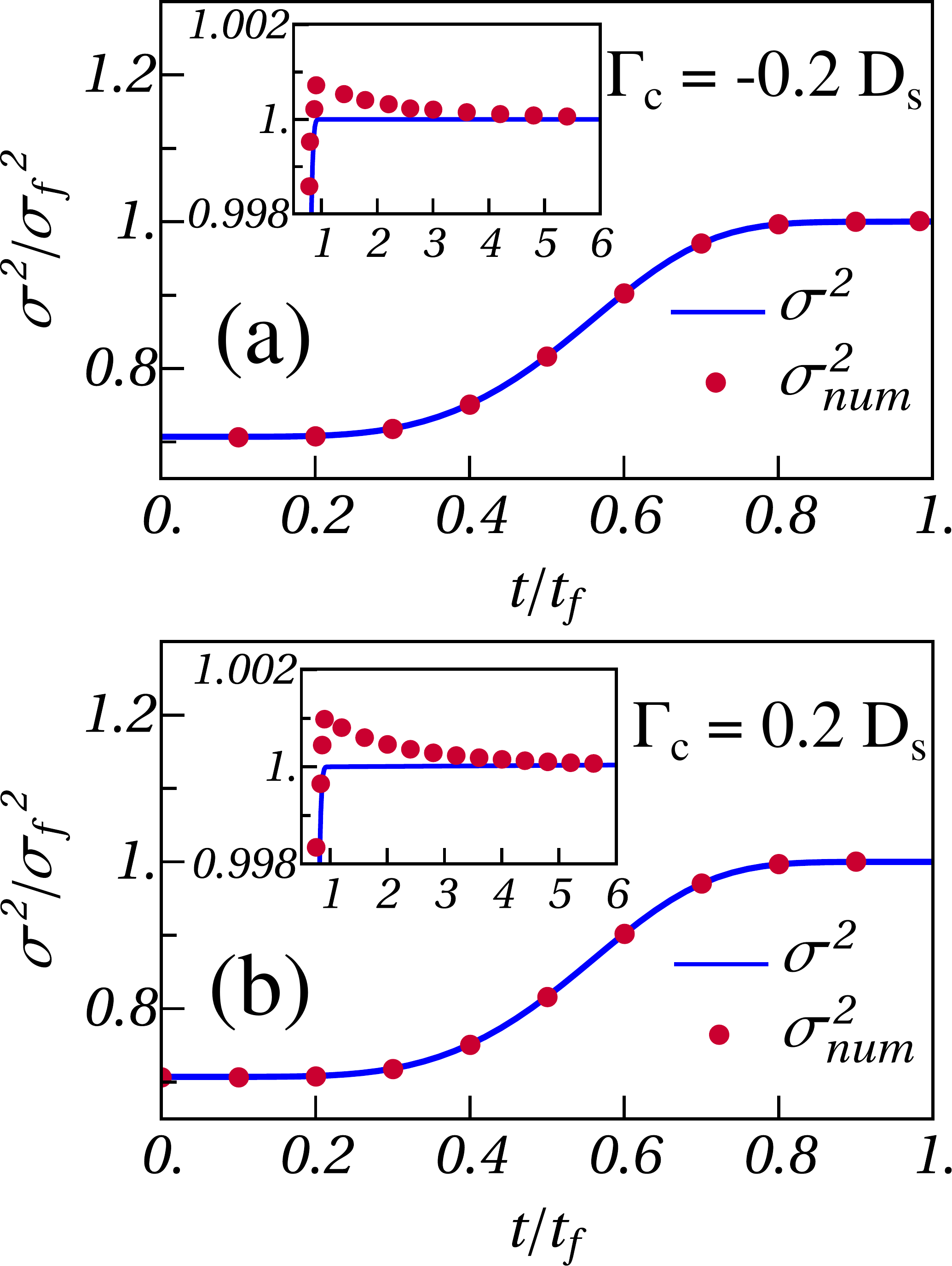}
\caption{\label{fig:var_1D_x4}
Non-Gaussian situation. Same as Fig. \ref{fig:pdf_1D_x4} for the variance of the distribution.
The inset emphasizes late time dynamics.
}
\end{figure}

\begin{figure}[!ht]
\centering \includegraphics[width=0.65\linewidth]{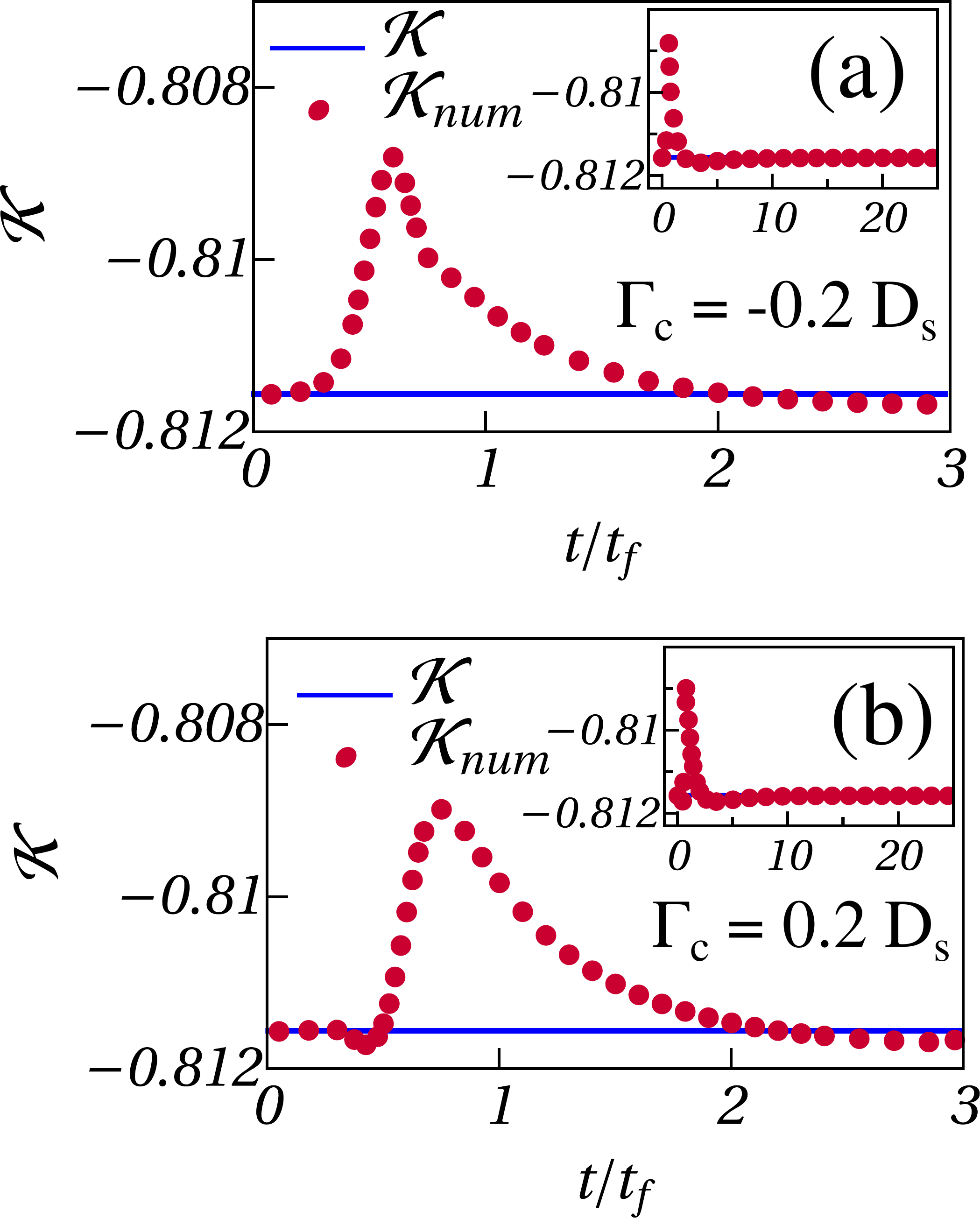}
\caption{\label{fig:kurtosis_1D_x4}
Same as Fig. \ref{fig:var_1D_x4}, for the reduced fourth moment of the  distribution (kurtosis).
}
\end{figure}

We next turn to the colloid density $\rho_{\mathrm{num}}(x,t)$, obtained from solving numerically the Fokker-Planck equation, Eq.~\eqref{eq:fokker}.
The results are shown in Fig.~\ref{fig:pdf_1D_x4}.
The density keeps the desired non-Gaussian distribution during the whole process, as shown at $t_{\mathrm{f}}$ in the Figure.
The agreement with the prescribed density is striking and outperforms what had been obtained
at Gaussian level. 
For times exceeding $t_{\mathrm{f}}$, a slight drift exists,
as displayed in the inset.
To ensure that the target expansion is realized, we calculate numerically the variance of the colloid $\sigma_{\mathrm{num}}^2(t) = \int_{-\infty}^{+\infty} x^2 \rho_{\mathrm{num}}(x,t)~dx$.
The results are plotted as a function of time for two phoretic mobilities $\Gamma_{\mathrm{c}} = \pm 0.2~D_{\mathrm{s}}$
in Fig.~\ref{fig:var_1D_x4}. 
The variance measured from the solution to the Fokker-Planck equation appears
very close to the target variance, $\sigma^2 = N/(2\sqrt{\alpha(t)})$.
The maximum error is observed for $t\simeq t_{\mathrm{f}}$ and 
is below $0.1$\%; the mismatch then quickly decays to zero (see the inset).
The plots confirm that we achieve a better control of non-Gaussian states than Gaussian ones (comparing Figs. \ref{fig:var_1D_x2} and \ref{fig:var_1D_x4}). The reason is that the
present treatment accounts for non harmonic contributions to the potential $U$,
while they are discarded in sections \ref{sec:ESE_DP_G} and \ref{sec:res_1D}.
Finally, for completeness, we evaluate the kurtosis of the colloidal density. This quantity, as shown in Fig.~\ref{fig:kurtosis_1D_x4}, is nearly constant and
changes in time only on a small scale,  such that ${\cal K} \simeq 0.8116 \pm 0.003$. This confirms that the colloidal density follows the target self-similar non-Gaussian distribution given by Eq.~\eqref{eq:x4_pdf} during the whole procedure. The kurtosis remains negative for both positive and negative mobilities, as a consequence of light tails in the target
distribution eq.~\eqref{eq:x4_pdf}. 

%


%
\section{Discussion and conclusion}
\label{sec:conclusion}
By a proper combination of optical and 
phoretic forcing, we have proposed a family 
of decompression protocols for manipulating
colloidal states. These protocols are free
of the difficulty that plagues an all optical 
method, namely the fact that the trap stiffness
in the vicinity of its center should become
negative, expelling colloids away,
in some time window. Here, diffusiophoretic forces
provide the required deconfining forces; they are monitored by an externally controlled 
salt buffer, that bathes the exterior of the 
system under study. In other words, we drive
our system by the (time-dependent) salt concentration $\phi(t)$ at its 
boundary, as it was realized in a series
of experimental works~\cite{palacci2010colloidal,palacci2012osmotic}.
It is a rather non trivial result that in doing so, we achieve the 
precise bulk forcing that is required to manipulate the colloidal
states under consideration. We take advantage here of the fast salt dynamics,
as compared to the slower colloidal response time. This is the key factor
that allows to decompress the colloidal state with a significant speedup compared
to the intrinsic response time.
We have shown that the
method is operational not only in a situation of harmonic confinement,
but also in a quartic external potential. 

After having derived analytically the appropriate drive $\phi(t)$, we have 
put our predictions to the test by first solving exactly  the salt diffusion 
equation from which the diffusiophoretic forces acting on the colloids follow;
these forces are then injected into the Fokker-Planck equation ruling colloidal
dynamics. An excellent agreement has been reported between the resulting
colloidal densities, and their (time-dependent) targets. For times larger
than the protocol duration $t_{\mathrm{f}}$, all evolution has to stop, and it is indeed
what has been observed. 
The colloidal dynamics is enslaved to that of salt through the Fokker-Planck equation, while salt itself obeys pure diffusion,
irrespective of colloidal arrangement. This is an approximation, valid for a low colloidal density. This is implicitly the limit in which we worked,
since all colloid-colloid interactions are neglected. An experimental way to 
achieve such a limit is to work with a single colloid, and gather statistics by repeating the experiment~\cite{martinez2016engineered,dago2020engineered}.
A simpler alternative deals with trapping a colloidal suspension, 
under the proviso that the maximal volume fraction remain small.

For simplicity, we have presented results in a one dimensional system, but we checked that the idea is also operational in two dimensions. Besides, it is possible to use a similar method to compress
the system rather than decompress it. While the added value of our protocol
for compression is less clear, since an all optical device may achieve a similar
result in a possibly more direct way, the present idea may nevertheless be used for 
speeding compression if the laser intensity cannot exceed some threshold, 
to avoid damaging the colloids.

The present contribution enriches the field of shortcut to adiabaticity protocols~\cite{guery2019shortcuts}, extending its underpinnings to non optical forcings in soft matter systems. While our method is approximate,
we have tested its accuracy. We also charted out the configurations (state points) that are amenable to our treatment, and accessible for a given acceleration factor $\cal A$.
Indeed, some state points (outside the green region in Fig. \ref{fig:phase_a2_1D_x2}) are not accessible. 
Yet, we did not attempt at optimality in any sense. 
There is thus room for improved protocols, that would for instance enlarge the
accessible region in parameter space (green regions in Fig.~
\ref{fig:phase_a2_1D_x2}), or aim at the best stability with respect to a
slight miscalibration of initial parameters,
through a careful choice of the --to a large extent arbitrary-- function $\alpha(t)$.

An open question deals with the possible challenges raised by the
experimental realization of the present ideas, starting with the precise
time control of the driving concentration $\phi(t)$, that may require some experimental
ingenuity. Besides, the question of optical 
stiffness anisotropy should be addressed.
For instance, we have assumed the coincidence of the 
point of zero optical force, with that of zero diffusiophoretic force. 
We briefly discuss in~\ref{app:misalignment} the consequences
of a misalignment, with the conclusion that the protocols is
relatively immune to a modest mismatch between the centers of optical 
and diffusiophoretic drivings.
Besides, one needs to change 
the salt concentration in a buffer that bathes from the exterior 
the system of interest, without perturbing hydrodynamically 
the colloidal bead's motion.
When addressing these questions, one should keep in mind that 
we made no attempt at optimizing the proposed protocols, which 
leaves a useful flexibility that can be taken advantage of
to minimize a given artifact. Optimization here also amounts to
selecting the relevant time-dependence for the positional variance,
through the choice of the function $\alpha(t)$.  We also 
note that {\it a priori} detrimental salt concentration fluctuations,
that could affect the diffusiophoretic drive,
can be reduced by increasing the initial uniform salt concentration 
$C_0$. It is also quite clear that implementing the non-Gaussian 
protocol leads to enhanced difficulties, since the joint control
of quadratic and quartic stiffness coefficients is required.

\subsection*{Acknowledgment}
Thanks are due to L. Bellon, L. Bocquet, S. Ciliberto, S. Dago, D. Gu\'ery-Odelin, I. Palaia, I. Pagonabarraga, C. Plata, A. Prados, and B. Rotenberg
for useful discussions.
The work was partially funded by Horizon 2020 program through 766972-FET-OPEN-NANOPHLOW.

\begin{appendix}

\section{Fast expansion beyond ESE}
\label{appen:expa_beyond_ESE}
As discussed in section~\ref{sec:DP}, fast decompression is possible making use of diffusiophoretic forces in an otherwise time-independent
optical potential, i.e., $\kappa_{\mathrm{op}} = \kappa_{\mathrm{i}}$. 
To obtain a non vanishing diffusiophoretic contribution to the confining stiffness, 
we need $\phi(t \geq	 0)=0$.
Without loss of generality, we consider again a one-dimensional system.
The analytical solution of the diffusion equation \eqref{eq:diffusion} in 1D
for a  system confined in $-L/2 \leq x \leq L/2$  reads 
\begin{widetext}
 \begin{eqnarray}
\label{C_1}
C(x,t) = &&\frac{4}{L} \sum_{n=0}^\infty (-1)^\textrm{n} \cos(\lambda_n x) e^{-\lambda_n^2 D_{\textrm{s}} t}
\left(\frac{1}{\lambda_n} + 
\lambda_n D_{\textrm{s}} \int_0^t e^{\lambda_n^2 D_{\textrm{s}} \omega}
\phi(\omega) d\omega\right)\nonumber\\
&& = \phi(t)-\frac{4}{L} \sum_{m=1}^\infty (-1)^m \frac{1}{D_{\textrm{s}}^m}
\sum_{n=0}^\infty \frac{(-1)^\textrm{n}}{\lambda_n^{2m+1}} \cos(\lambda_n x) 
\left(\frac{d^{m}\phi(t)}{dt^{m}} + e^{-\lambda_n^2 D_{\textrm{s}} t} \frac{d^{m}\phi(t)}{dt^{m}}|_{t=0} \right),
\end{eqnarray}
\end{widetext}
where $\lambda_n = (2n+1)\pi/L$
and $\phi(t)=C(\pm L/2,t)$ is the boundary 
concentration, playing the role of the driving field in our analysis 
and monitored by the experimentalists. Considering $\phi(t \geq	 0)=0$,
we have for $t\gg  \tau_{\mathrm{s}} \equiv L^2/ D_{\mathrm{s}}$,
$ C(x,t) \propto \cos(\pi x/L) \exp(-\pi^2 D_{\mathrm{s}} t/L^2) $
which features factorized $x$ and $t$ dependencies.
The diffusiophoretic potential follows as 
\begin{equation}
U_{\mathrm{DP}}(x) \,=\, -\gamma \Gamma_{\mathrm{c}} \, 
\log\left(\cos\left(\frac{\pi x}{L}\right)\right).
\label{eq:UDP1dfull}
\end{equation}
Close enough to $x=0$, this can be Taylor expanded,
and $U_{\mathrm{DP}}$ therefore emulates harmonic trapping with stiffness 
\begin{equation}
\kappa_{\mathrm{DP}}
= 
\gamma \Gamma_{\mathrm{c}} \frac{\pi^2}{L^2},
\end{equation}
which only depends on the system size  $L$.
The time-scale required to reach such a state is
$ 
\tau_{\mathrm{s}}$. 

The potential energy close to the trap center, with optical and diffusiophoretic terms, is then approximated by $U \simeq \frac{1}{2} (\kappa_{\mathrm{op}} + \gamma \Gamma_{\mathrm{c}} \pi^2 / L^2)) x^2$.
The density
$\rho(x,t)$ 
evolves according to the Fokker-Planck equation~\eqref{eq:fokker}
which for the one dimensional system, at long times $t\gg \tau_{\mathrm{s}}$, and near the center 
$|x| \ll L$, reduces to 
\begin{equation}
 \partial_t \rho({\bf r},t) \simeq \partial_x \left(D_{\mathrm{c}} \partial_x \rho(x,t) 
 +   \left(\frac{\kappa_{\mathrm{op}}}{\gamma} +\Gamma_{\mathrm{c}} \frac{\pi^2 }{L^2}\right)x \rho(x,t)\right).
\end{equation}
With an initial Gaussian density $\rho(x,0)$ (variance $\sigma_{\mathrm{i}}^2$), 
$\rho(x,t)$ remains Gaussian in a harmonic trap, and as long as $|x| \ll L$
the solution reads
\begin{equation}
\rho(x,t) = \sqrt{\frac{g(t)}{2 \pi \sigma_{\mathrm{f}}^2}} e^{- g(t) \frac{x^2}{2\sigma_{\mathrm{f}}^2}},
\end{equation}
where $g(t) = \kappa_{\mathrm{i}}/\left((\kappa_{\mathrm{i}}+(\kappa_{\mathrm{f}}-\kappa_{\mathrm{i}})
e^{-2 D_{\mathrm{s}} t/\sigma_{\mathrm{f}}^2}\right)$.
Particularly, we have
\begin{equation}
\rho(x,t\rightarrow \infty) = \sqrt{\frac{1}{2\pi\sigma_{\mathrm{f}}^2}} e^{-\frac{x^2}{2\sigma_{\mathrm{f}}^2} },
\end{equation}
with $\sigma_{\mathrm{f}}^{2} = k_{\mathrm{B}} T/\kappa_{\mathrm{f}}$. Linearization 
of relation \eqref{eq:UDP1dfull}
is thus justified as long as 
$\sigma_{\mathrm{f}} \ll L$.

We conclude here that we have realized a fast 
decompression, on a time scale $\tau_{\mathrm{s}}$,
where nevertheless the final position variance
is determined by $L$, the system size,
through the effective stiffness 
\begin{equation}
\kappa_{\mathrm{f}} \,=\, \kappa_{\mathrm{op}} + \kappa_{\mathrm{DP}} 
\,=\, \kappa_{\mathrm{op}} +
\gamma \,\Gamma_{\mathrm{c}} \,\frac{\pi^2}{L^2}.
\end{equation}
Moreover, we can see that 
$\kappa_{\mathrm{DP}}$ takes negative values only for $\Gamma_{\mathrm{c}} < 0$, which would not allow to decompress a system with $\Gamma_{\mathrm{c}} > 0$.
Therefore, the present method lacks control and generality,
and another route has been explored in the main text.

\section{Constraint on $C(0,t)$}
\label{appen:C0}
In this appendix, we present the calculations behind Eq.~\eqref{eq:den_cond}.
%
The assumptions of isotropicity and analyticity of $C({\bf r},t)$ imply $\partial^j_{r} C|_{r=0} =0$ for $j$ odd, so that 
expansion of $C({\bf r},t)$ at ${r =0}$ gives
\begin{eqnarray}
C({\bf r},t) = C(0,t) &+& \partial^2_{r} C({\bf r},t)|_{r =0} \frac{r^2}{2} \nonumber\\
&+& \partial^4_{r} C({\bf r},t)|_{r =0} \frac{r^4}{4!}
 + {\cal O}(r^6),
\end{eqnarray}
which reveals that 
$\frac{1}{r} \partial_{r} C|_{r=0} = \partial^2_{r} C|_{r=0}$ 
(although $\partial_{r} C|_{r=0} =0$).
Then, for $\nabla^2 C$ in $n-$dimensional system at $r=0$ we have
\begin{eqnarray}
 &&\nabla^2 C|_{r=0} = \partial^2_{r} C({\bf r},t)|_{r=0} 
 + \frac{n-1}{r} \partial_{r} C({\bf r},t)|_{r=0}\nonumber\\
 && = n~\partial^2_{r} C({\bf r},t)|_{r=0}.
\end{eqnarray}
Substituting this relation into the diffusion equation  \eqref{eq:diffusion} gives
\begin{equation}
\partial_t C({\bf r},t)|_{r=0} = D_{\mathrm{s}}~ n~\partial^2_{r} C|_{r=0}.
\end{equation}
Thus, we have
\begin{eqnarray}
 &&\partial^2_{r} \ln C({\bf r},t)|_{r=0} \nonumber\\
 &&= \frac{1}{C}\partial^2_{r} C({\bf r},t)\biggl|_{r=0} - \frac{1}{C^2}
\left(\frac{}{}\partial_{r} C({\bf r},t)\right)^2\biggl|_{r=0} \nonumber\\
&& = \frac{1}{n~D_{\mathrm{s}}} \frac{1}{C} \partial_{t} C({\bf r},t)|_{r=0}
= \frac{1}{n~D_{\mathrm{s}}} \partial_t \ln C(0,t).
\end{eqnarray}
Combining this equation with Eqs.~\eqref{eq:kappa_DP} and \eqref{eq:kappa}, we obtain 
a differential equation for the concentration at the center $C(0,t)$ as
\begin{equation}
\label{eq:den_cond_appen}
 -\frac{\gamma \Gamma_{\mathrm{c}}}{n~D_{\mathrm{s}}}  \partial_t \ln C(0,t)
  = \frac{\dot{\alpha}(t)}{2 \alpha(t)} \gamma + 2 k_{\mathrm{B}} T \alpha(t) - \kappa_{\mathrm{op}}(t)
\end{equation}

\section{Results for $\alpha^{(1)}(t)$}
\label{appen:res_alpha1}
\begin{figure}[htb]
\centering \includegraphics[width=1\linewidth]{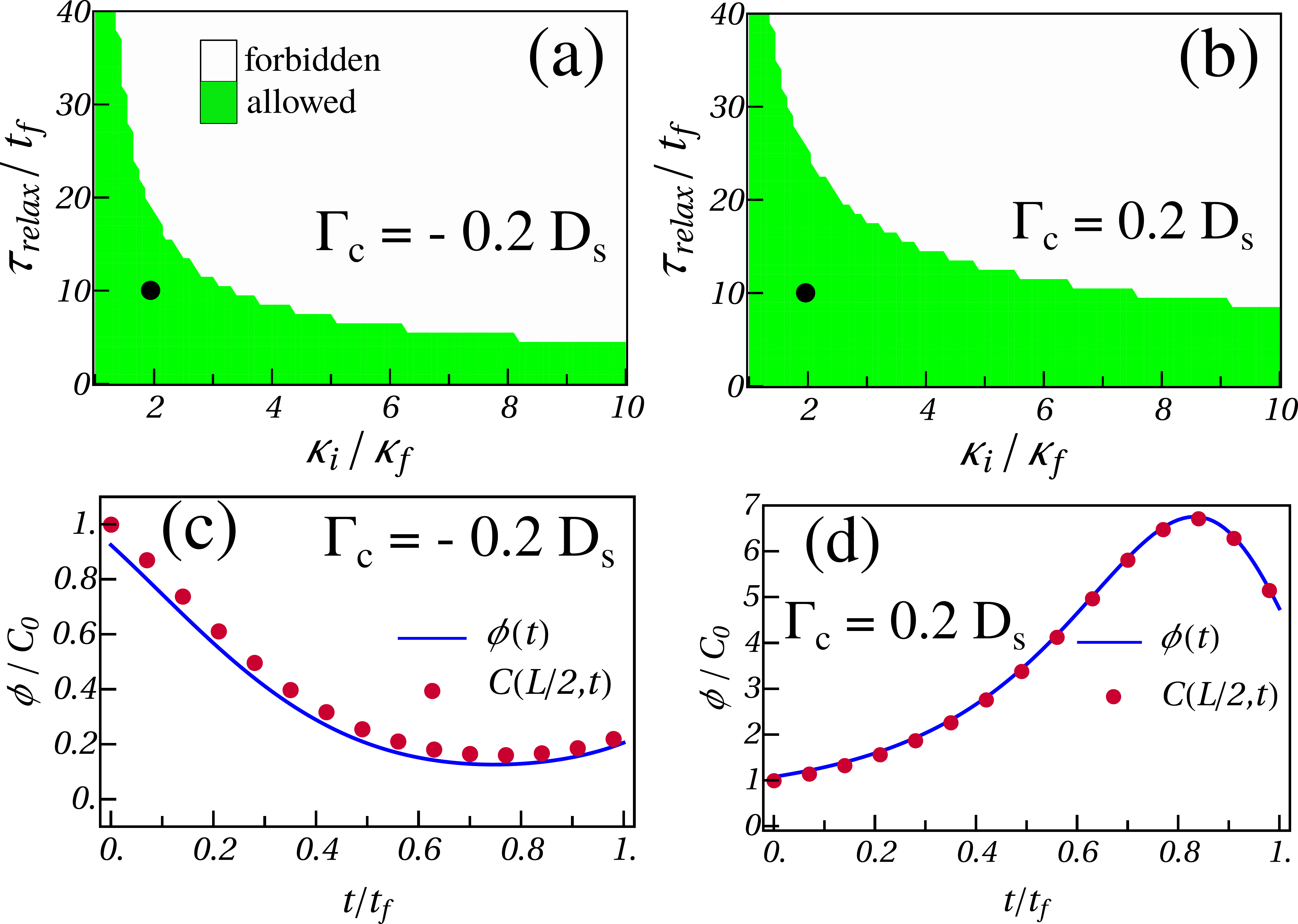}
\caption{\label{fig:phase_a1_1D_x2} 
Same as Fig.~\ref{fig:phase_a2_1D_x2} with $\alpha(t) = \alpha^{(1)}(t)$ 
rather than $\alpha(t) = \alpha^{(2)}(t)$. All parameters are the same, while we also took $\kappa_{\mathrm{op}}(t)=2k_{\mathrm{B}} T\alpha(t)$. The state points corresponding to panels c) and d) 
are shown with a black dot in panels a) and b). 
}
\end{figure}

In the main text, we presented results for the situation where the inverse
variance of colloidal positions, encoded in the function $\alpha(t)$, is given
by $\alpha^{(2)}$ in Eq.~\eqref{eq:alpha2}. We work out here the method with another choice ($s=t/t_{\mathrm{f}}$),
\begin{equation}
    \alpha(t)=\alpha^{(1)}(t) 
= \frac{1}{2 k_{\mathrm{B}} T} \left(\frac{}{}\kappa_{\mathrm{i}} + (\kappa_{\mathrm{f}}-\kappa_{\mathrm{i}}) (3s^2-2s^3)\right),
\end{equation}
that highlights the importance of the steadiness of $\alpha$, essentially near the 
end point $t=t_{\mathrm{f}}$. Here also, we suppose $\kappa_{\mathrm{op}}(t) = 2 k_{\mathrm{B}} T\alpha(t)$. The resulting phase portrait, the counterpart of Fig.~
\ref{fig:phase_a2_1D_x2}, is shown in Fig.~\ref{fig:phase_a1_1D_x2}.

\begin{figure}[htbp]
\centering \includegraphics[width=0.65\linewidth]{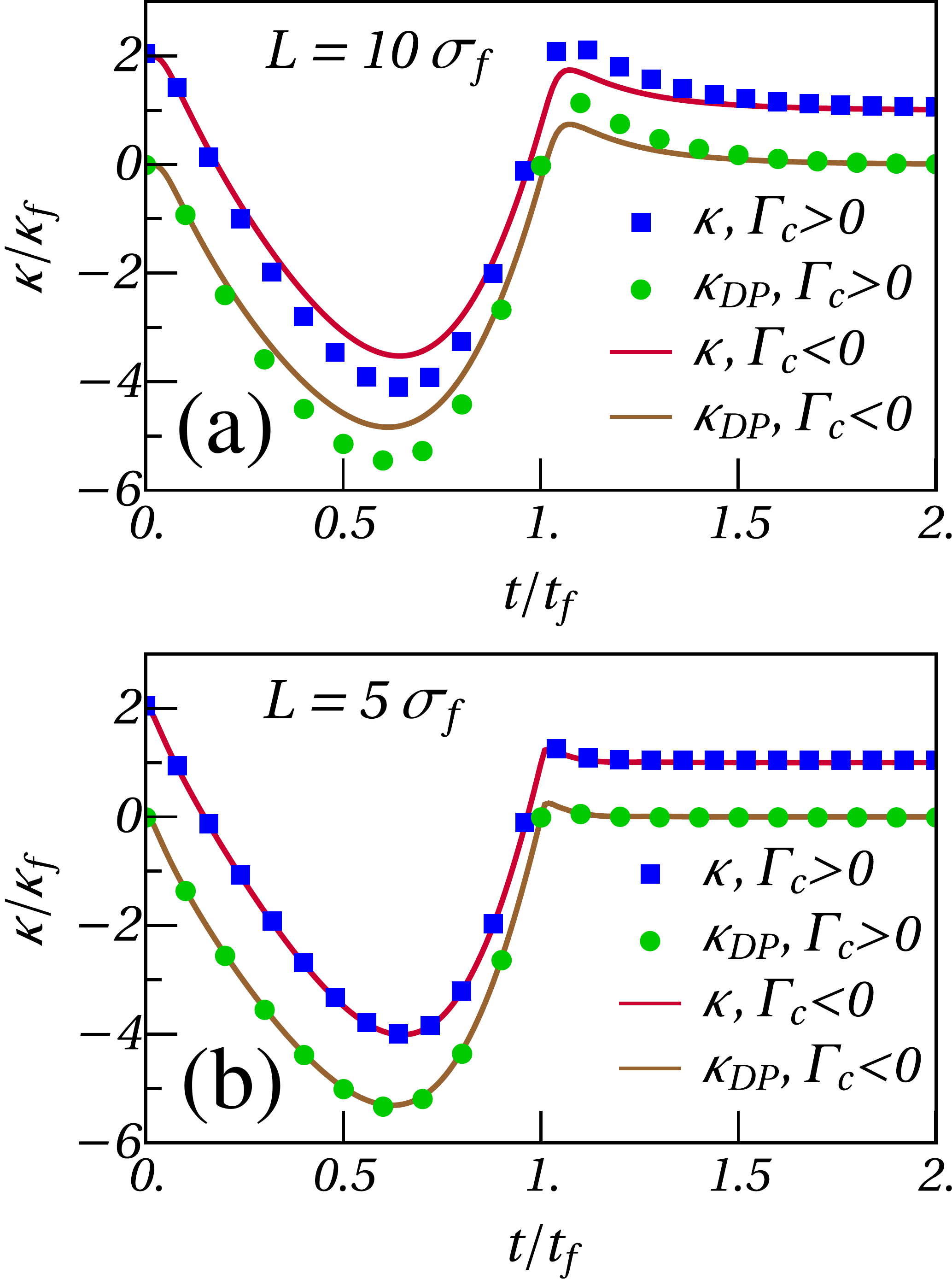}
\caption{\label{fig:kappa1_all} 
The stiffnesses $\kappa(t)$ and $\kappa_{\mathrm{DP}}(t)$, as a function of time for parameters 
${\cal A} = 10,~\kappa_{\mathrm{i}}=2 \kappa_{\mathrm{f}}$, $D_{\mathrm{c}} = 0.002D_{\mathrm{s}}$, 
$\kappa_{\mathrm{op}}(t)= 2 k_{\mathrm{B}} T\alpha^{(1)}(t)$ and for two phoretic mobilities:
$\Gamma_{\mathrm{c}} = \pm 0.2~D_{\mathrm{s}}$; 
here, $\alpha(t)=\alpha^{(1)}(t)$.
The two graphs are for different sizes
(a) $L=10 \sigma_{\mathrm{f}}$ and (b) $L=5 \sigma_{\mathrm{f}}$. 
Note the undesired time-dependence of the quantities 
displayed, for $t> t_{\mathrm{f}}$.
The data collapse in panel 
(b) is indicative of a better accuracy of the protocol, as confirmed 
by the near constancy of the stiffnesses for $t> t_{\mathrm{f}}$.
}
\end{figure}

Two comments are in order. First and for $\Gamma_{\mathrm{c}}<0$, the $\alpha^{(1)}$ protocol extends the 
reach of the method as compared to the $\alpha^{(2)}$ protocol. The reverse 
conclusion holds for $\Gamma_{\mathrm{c}}>0$. Second, and more relevant,
the accuracy does not fare favorably to that shown in panel (c) of 
Fig.~\ref{fig:phase_a2_1D_x2}. With positive 
mobilities (panel d)) we observe a better 
collapse of the analytically computed $\phi(t)$ shown 
with the line, and the symbols resulting from the numerical resolution 
of the salt diffusion equation. In panel c), the lack of accuracy is in particular 
already visible at $t=0$, where $\phi$ slightly departs from $C_0$,
the salt buffer concentration. This is a consequence of our Taylor expansion 
truncation, not only to get $\phi(t)$, but also the potential 
$U$, that may feature non-harmonic contributions. 
These shortcomings are made more apparent when looking at the dynamics of the 
resulting stiffness $\kappa$, or $\kappa_{\mathrm{DP}}$ as in Fig.~\ref{fig:kappa1_all}. It is seen 
that for $t>t_{\mathrm{f}}$, these quantities are not steady.
Besides, the effect of the sign of the mobility $\Gamma_{\mathrm{c}}$, at variance
with the programmed solution (see Fig.~\ref{fig:kappa_all_1D_x2}),
is another illustration of the imperfections at stake here. 
As expected and observed in comparing the two graphs 
in Fig \ref{fig:kappa1_all}, 
the protocol works better when applied to a smaller system.

\section{Non-Gaussian features of would-be Gaussian states}
\label{appen:nonG_G}
\begin{figure}[!ht]
\centering \includegraphics[width=0.65\linewidth]{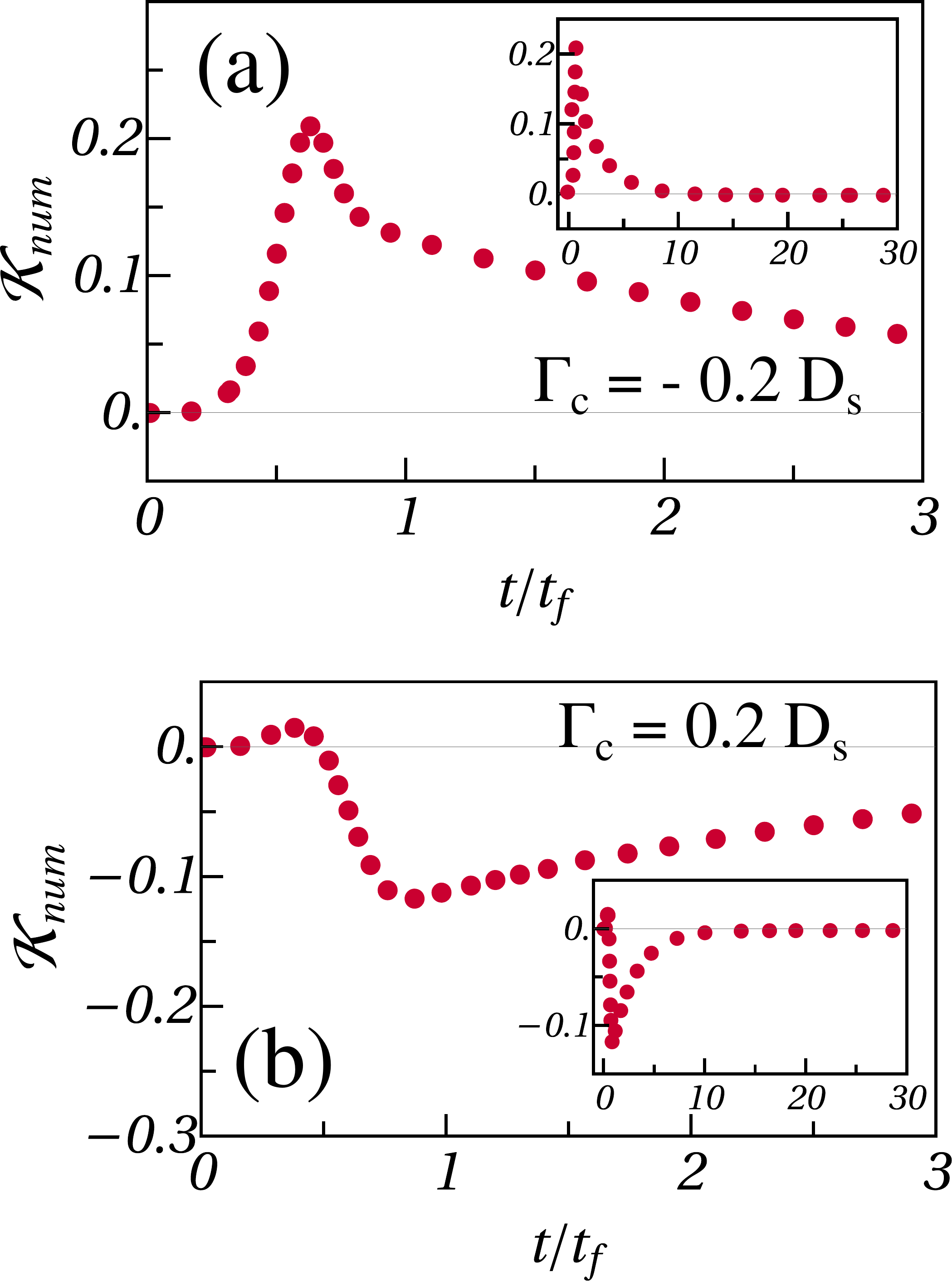}
\caption{\label{fig:kurtosis_1D_x2_A}
Kurtosis of the colloidal position distribution as a function of time for the target Gaussian states. The panels correspond to (a) $\Gamma_{\mathrm{c}} = -0.2~D_{\mathrm{s}}$ and (b) $\Gamma_{\mathrm{c}} = 0.2~D_{\mathrm{s}}$ for $\alpha(t)$ given by Eq.~\eqref{eq:alpha2} and the same parameters as in Fig.~\ref{fig:phase_a2_1D_x2}.
The insets show the dynamics for long times.
}
\end{figure}

In sections~\ref{sec:ESE_DP_G} and~\ref{sec:res_1D}, we have supposed that the colloid density remains Gaussian during the protocol. In this appendix, we question the validity of this assumption.
To this end, we calculate numerically the kurtosis, which measures the non-Gaussianity of the colloidal density.
While the skewness remains zero by symmetry, the kurtosis does not strictly vanish, at variance with an exact Gaussian density. The numerically calculated result ${\cal K}_{\mathrm{num}}(t) = \int_{-\infty}^{+\infty} (x-\langle x \rangle)^4 \rho_{\mathrm{num}}(x,t)~dx/\sigma^4 - 3$ is shown in Fig.~\ref{fig:kurtosis_1D_x2_A}. This non-zero kurtosis originates from the quartic term in the total potential energy and 
leads the colloid density to deviate from Gaussian distribution. However, the maximum kurtosis is around $0.2$ which is rather small.

\begin{figure}[!ht]
\centering \includegraphics[width=0.75\linewidth]{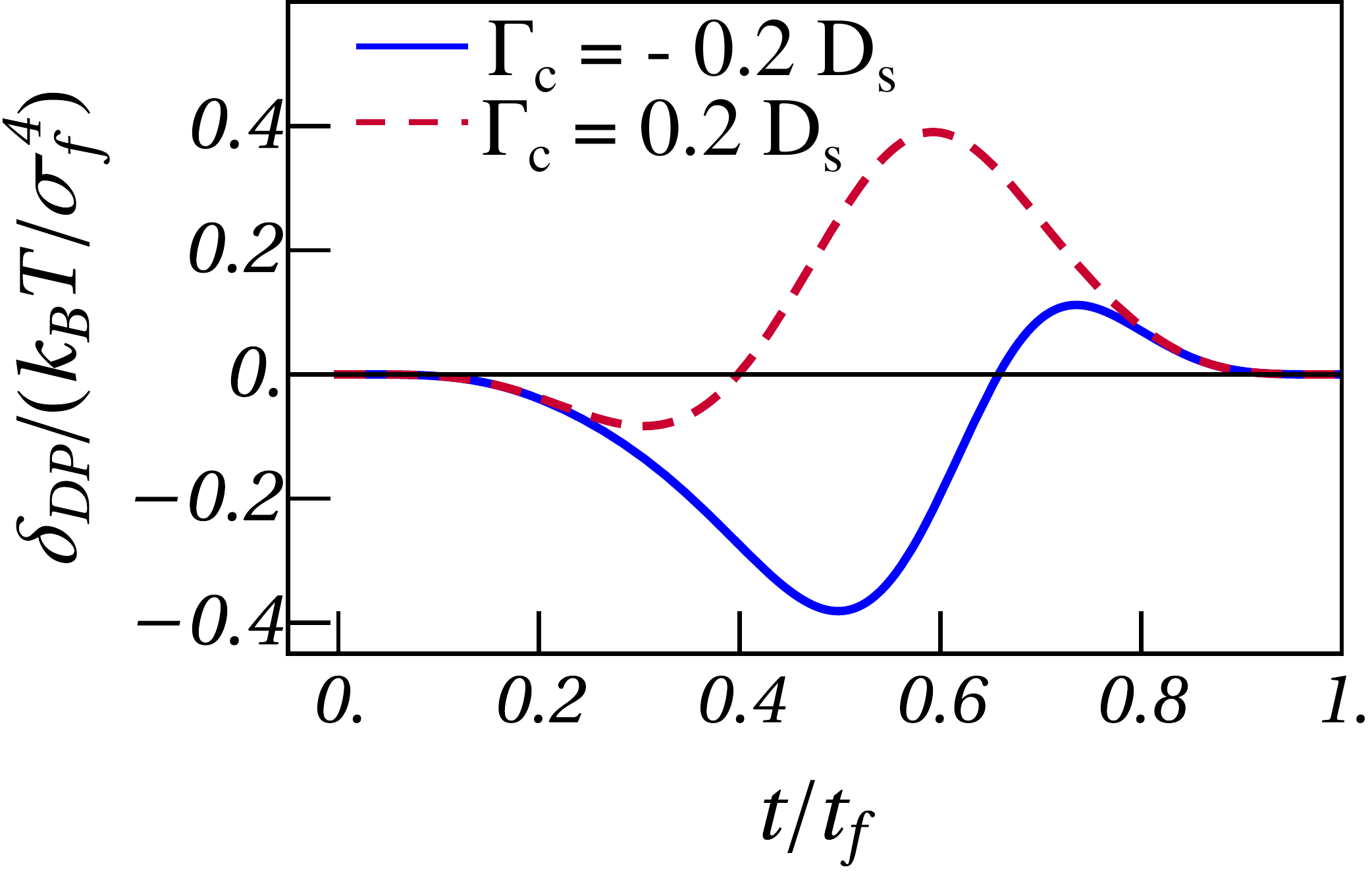}
\caption{\label{fig:delta_DP_G}
Quartic stiffness coefficient $\delta_{\mathrm{DP}}$ as a function of time for the Gaussian states for two phoretic mobilities. The panels correspond to the same parameters as in Fig.~\ref{fig:phase_a2_1D_x2} and $\alpha(t)$ given by Eq.~\eqref{eq:alpha2}.
}
\end{figure}

Furthermore, the kurtosis depends on the phoretic mobility.
This is explained by considering the quartic term in the potential energy 
($\delta_{\mathrm{DP}} \,x^4/4)$, where $\delta_{\mathrm{DP}}(t)= - \frac{1}{6} \gamma \Gamma_{\mathrm{c}} \partial^4_{x}\ln C(x,t)|_{x=0}$. As shown in Fig.~\ref{fig:delta_DP_G}, unlike the quadratic term, $\delta_{\mathrm{DP}}(t)$ depends on the phoretic mobility. For $\Gamma_{\mathrm{c}} < 0$, $\delta_{\mathrm{DP}}$ is mostly negative. This pushes the colloids further towards the exterior (enhanced expulsive contribution).
Thus, there will be an overpopulation at larger
distances and a positive kurtosis. For $\Gamma_{\mathrm{c}} > 0$, the argument is reversed (mostly positive $\delta_{\mathrm{DP}}$, mostly negative kurtosis). 

\section{Consequences of misaligning of the optical and diffusiophoretic forces}
\label{app:misalignment}

\begin{figure*}[htbp]
\centering \includegraphics[width=0.7\linewidth]{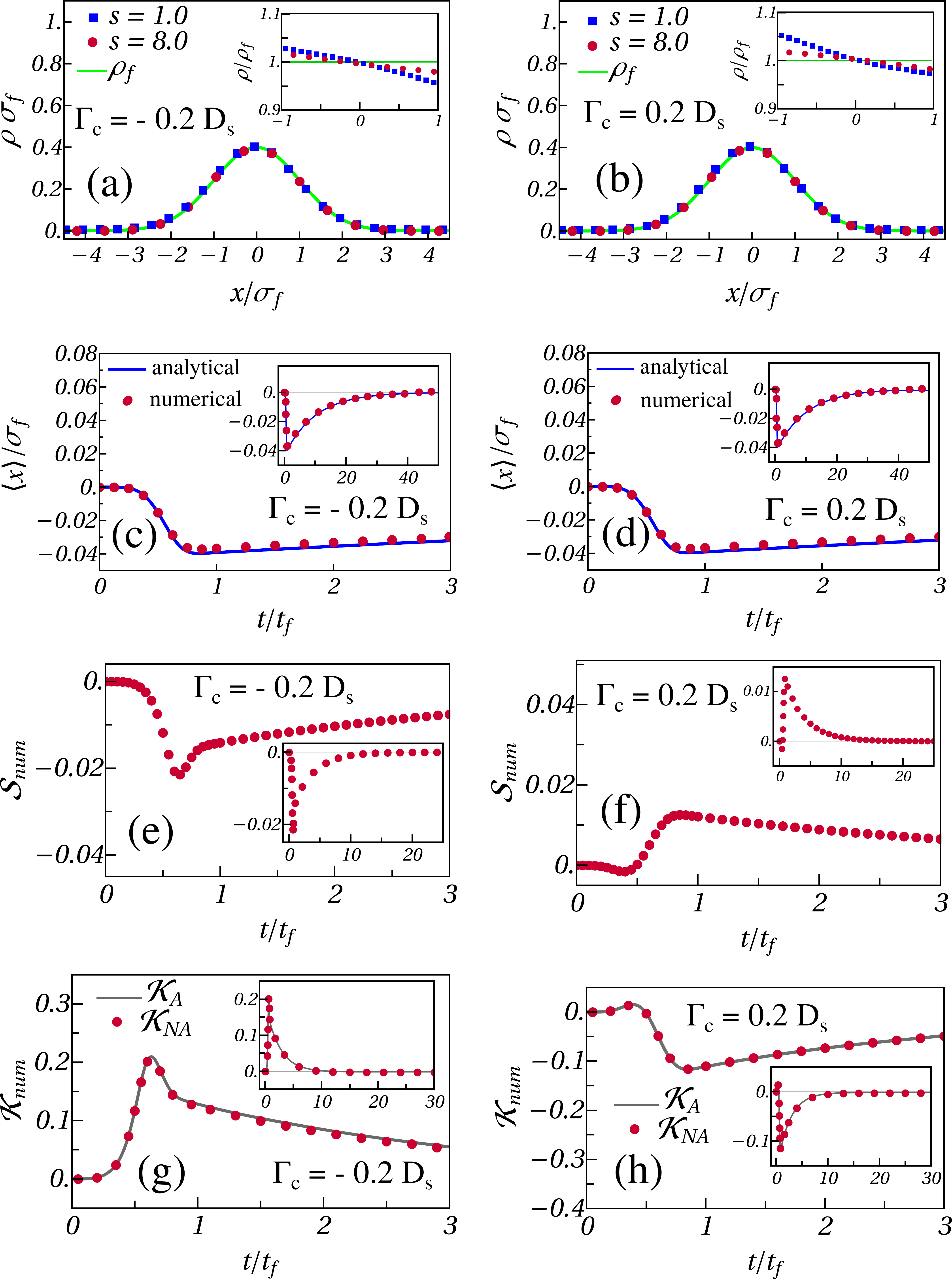}
\caption{\label{fig:non-a} 
Results for non-aligned optical and diffusiophoresis forces: (a--b) colloid density (c--d), mean value of colloid position $\langle x(t) \rangle$, (e--f) skewness, and (g--h) kurtosis of the colloid distribution, for the same parameters as Fig.~\ref{fig:phase_a2_1D_x2} and $x_0=0.1 \sigma_{\mathrm{f}}$. For the kurtosis, the subscripts
{\it A} and {\it NA} refer to the aligned and non-aligned situations.
}
\end{figure*}

\begin{figure*}[htbp]
\centering \includegraphics[width=0.5\linewidth]{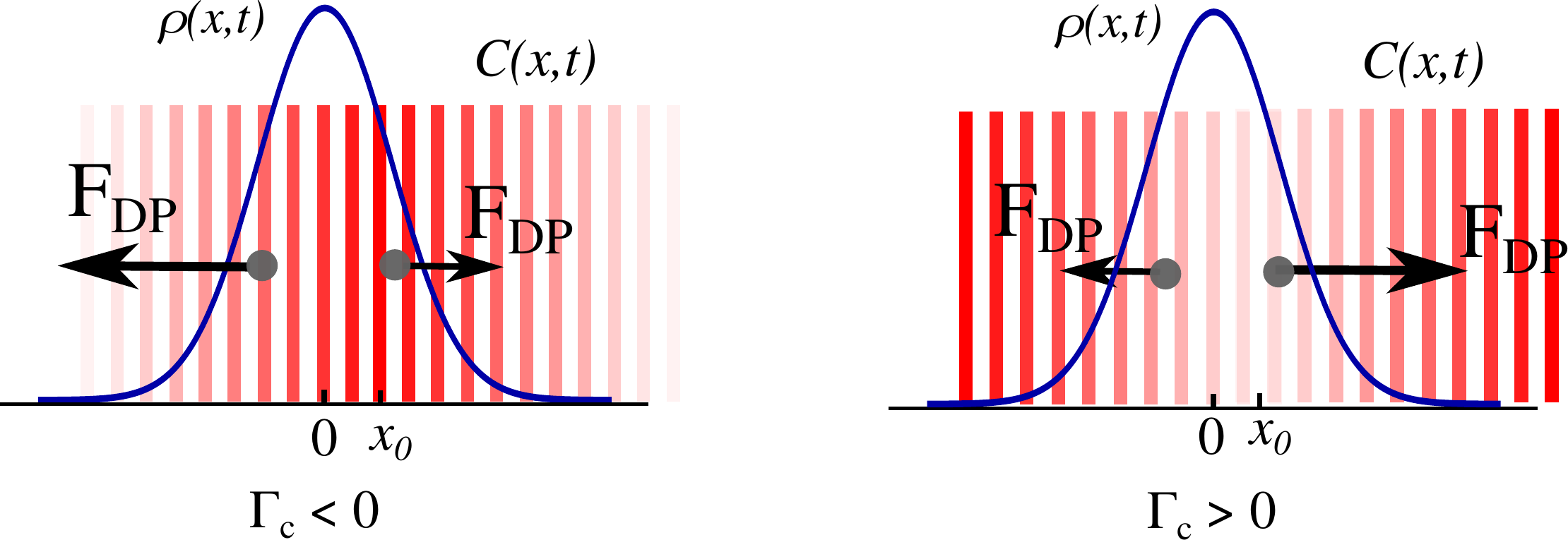}
\caption{Force bias at the origin of the non-vanishing skewness 
for optical and chemical forcing non-alignment. Optical forces are 
symmetric around $x=0$ while diffusiophoretic forces are centered around
$x=x_0>0$, which leads to the salt density visualized by the red shade.
\label{fig:non-a-sketch} 
}
\end{figure*}

We address here a possible experimental limitation, coming from the fact that 
it may prove difficult to have a perfect coincidence between the minimum 
of the optical potential created, and the point of symmetry of the salt driving,
where the diffusiophoretic forces vanish. We therefore assume that
the optical and diffusiophoretic forces are centered at $x=0$ and $x=x_0 \neq 0$, respectively. Then, the total potential energy is given by
\begin{eqnarray}
U &=& \frac{1}{2} \kappa_{op}(t) x^2 -\gamma \Gamma_c \ln C(x-x_0,t)\nonumber\\
&\simeq& \frac{1}{2} \kappa(t) x^2 - \kappa_{\mathrm{DP}}(t)\ x_0\ x,
\end{eqnarray}
where $\kappa(t) = \kappa_{\mathrm{op}} + \kappa_{\mathrm{DP}}$, as in the main text, with
\begin{equation}
\kappa_{\mathrm{DP}}(t) = -\gamma \Gamma_c \partial_{x,x}^2 \ln C(x,t)|_{x=x_0}.
\end{equation}
This results in a time-dependent trap center, located at
$x= \kappa_{\mathrm{DP}}(t) \,x_0/\kappa(t)$. 
The assumptions of isotropicity and analyticity of $C(x,t)$ still hold with $\partial^j_{r} C|_{x=x_0} =0$ for $j$ odd. This potential leads to a Gaussian colloidal position distribution with a time-dependent mean value $\langle x(t) \rangle$:
\begin{equation}
\rho(x,t) = \sqrt{\frac{\alpha(t)}{\pi}} e^{-\alpha(t) 
\left[ x-\langle x(t) \rangle \right]^2}.
\end{equation}
Substituting $\rho(x,t)$ into the Fokker-Planck equation, Eq.~\eqref{eq:fokker}, in addition to the rule given by Eq~\eqref{eq:kappa} for $\kappa(t)$, we obtain a differential equation governing $\langle x \rangle$ as
\begin{equation}
\label{eq:meanx_D}
\gamma\ \partial_{t} \langle x(t) \rangle = \kappa_{\mathrm{DP}}(t) x_0 - \kappa(t) \langle x(t) \rangle.
\end{equation}
Solving for $ \langle x(t) \rangle$ with initial condition $\langle x(t=0) \rangle =0 $ yields
\begin{equation}
\label{eq:meanx}
\langle x(t) \rangle = \int_0^t d\tau~ \frac{\kappa_{\mathrm{DP}}(\tau) x_0}{\gamma}
e^{-\int_0^\tau dt'~ \kappa(t')/\gamma     }.
\end{equation}
Since non-aligning does not change the evolution of $\kappa(t)$ in Eq.\eqref{eq:kappa}, the protocol remains the same as presented in the main text for the aligned case with  
$\kappa_{\mathrm{op}}(t) = 2 k_{\mathrm{B}} T \alpha(t)$ 
and $\alpha(t)$ given in Eq.~\eqref{eq:2alpha}.
Hence, the variance does not change and corresponds to Fig.~\ref{fig:var_1D_x2}. 
We consider the changes of other moments.
The colloidal density $\rho_{\mathrm{num}}$ is obtained by solving numerically the Fokker-Planck equation Eq.~\eqref{eq:fokker}. The result is shown in Fig.~\ref{fig:non-a}(a--b) for the same parameters as in Fig.~\ref{fig:phase_a2_1D_x2} and $x_0 = 0.1 \sigma_{\mathrm{f}}$. It is seen, especially in the inset plots, that the other moments, i.e., the mean value, skewness and kurtosis, are subject to changes. 
Then, we evaluate $\langle x \rangle$ either from
\eqref{eq:meanx_D}, or by calculating numerically $\langle x \rangle_{\mathrm{num}} = \int_{-\infty}^{\infty} x(t) \rho_{\mathrm{num}} ~dx$. 
As Fig.~\ref{fig:non-a}(c--d) and also the inset plots in panels (a--b) show, $\langle x \rangle$ starts from zero at initial time and reaches a maximum value. According to Eq. \eqref{eq:meanx}, the maximum value of deviation from initial trap center is given by 
$\langle  x \rangle_{\mathrm{max}} = \kappa_{\mathrm{DP}}(t^*) x_0/\kappa(t^*)$ at $t^*$, when both the time dependent trap center $x_{\mathrm{trap}}$ ($\nabla U(x_{\mathrm{trap}},t)=0$) and the mean value overlap.
At the end of the protocol when $t=t_{\mathrm{f}}$, $\langle x \rangle$ differs slightly from the target trap center at $x=0$. This deviation is as a result of the inertia of the colloid: although the ions concentration is uniform at $t=t_{\mathrm{f}}$ and the diffusiophoretic force vanishes, the colloid is on average
slightly away from the optical potential minimum at $x=0$, which results in a further drift.

To investigate if the colloid density remains Gaussian, we calculate the skewness ${\cal S}_{\mathrm{num}} = (x-\langle x \rangle)^3/\sigma^3$ and the kurtosis.
The skewness is shown in Fig.~\ref{fig:non-a}(e--f). Non-alignment leads to a non-vanishing value, which is expected because of 
the mismatch between the center of diffusiophoretic forcing at $x_0$ and the optical center at $x=0$. Moreover, the skewness depends on the phoretic mobility such that it is positive for $\Gamma_{\mathrm{c}}>0$ and negative for $\Gamma_{\mathrm{c}}<0$. The sketch in
Fig. \ref{fig:non-a-sketch} explains that a negative skewness ensues
for a positive $x_0$ when $\Gamma_c<0$, while the converse holds
for $\Gamma_c>0$.
Finally, as Fig.~\ref{fig:non-a}(g--h) shows and as expected, non-alignment does not affect the kurtosis. The kurtosis here (red symbols) is almost the same as the one for the aligned case (shown by a solid gray curve). 
%

\end{appendix}
%

\begin{thebibliography}{61}
\bibitem [{not()}]{comment203}%
  \BibitemOpen
  \href@noop {} {}\bibinfo {note} {In practice, this limit
  does not seem too much demanding. Indeed, the ratio
  $D_{\mathrm{s}}/D_{\mathrm{c}}$ is on the order of $10^3$ for a micron-sized
  colloid. Thus, $R/\sigma_{\mathrm{f}}$ should not exceed 30. Since
  $\sigma_\mathrm{f}$ is a measure of the spread of colloidal positions, $R$
  should be larger than a few $\sigma_\mathrm{f}$ in order not to interfere
  with the colloidal displacements, but less than $30
  \sigma_\mathrm{f}$.}\BibitemShut {Stop}%
\end{thebibliography}

%

\end{document}